\documentclass[aps,prx,twocolumn,superscriptaddress,longbibliography,10pt]{revtex4-2}
\usepackage{graphicx} 
\usepackage{amsmath}
\usepackage{amssymb}
\usepackage{soul}
\usepackage[table]{xcolor}
\usepackage[colorlinks=true, linkcolor=black, citecolor=blue, urlcolor=blue]{hyperref}
\usepackage[utf8]{inputenc}
\usepackage[T1]{fontenc}
\usepackage{tikz}
\usepackage[braket]{qcircuit}
\usepackage{pgffor}
\usepackage{orcidlink}
\usepackage{array,multirow}
\usepackage{makecell}
\usepackage{siunitx}
\usepackage{booktabs}
\usepackage{tikz}
\usepackage{mathtools}
\usepackage[normalem]{ulem}
\usepackage[version=4,arrows=pgf-filled,
textfontname=sffamily,
mathfontname=mathsf]{mhchem}

\usepackage{pifont}
\definecolor{gold}{rgb}{0.83, 0.69, 0.22}

\newcommand{\Mark}[1]{#1}

\begin{document}

\title{\texorpdfstring{Faster quantum chemistry simulations on a quantum computer \\with improved tensor factorization and active volume compilation}{Faster quantum chemistry simulations on a quantum computer}}

\newcommand{\BI}{\affiliation{
Quantum Lab, Boehringer Ingelheim, 55218 Ingelheim am Rhein, Germany}}

\newcommand{\BIMedChem}{\affiliation{Medicinal Chemistry, Boehringer Ingelheim Pharma GmbH \& Co. KG, 88397 Biberach, Germany}}

\newcommand{\PSIQ}{\affiliation{PsiQuantum, 700 Hansen Way, Palo Alto, CA 94304, USA}}

\author{Athena Caesura}
\PSIQ

\author{Cristian L.~Cortes}
\PSIQ

\author{William Pol}
\PSIQ

\author{Sukin Sim}
\PSIQ

\author{Mark Steudtner}
\email{msteudtner@psiquantum.com}
\PSIQ

\author{Gian-Luca R.~Anselmetti\orcidlink{0000-0002-8073-3567}}
\BI

\author{Matthias Degroote\orcidlink{0000-0002-8850-7708}}
\email{matthias.degroote@boehringer-ingelheim.com}
\BI

\author{Nikolaj Moll\orcidlink{0000-0001-5645-4667}}
\BI

\author{Raffaele Santagati\orcidlink{0000-0001-9645-0580}}
\BI

\author{Michael Streif\orcidlink{0000-0002-7509-4748}}
\BI

\author{Christofer S.~Tautermann\orcidlink{0000-0002-6935-6940}}
\BIMedChem

\begin{abstract}
Electronic structure calculations of molecular systems are among the most promising applications for fault-tolerant quantum computing (FTQC) in quantum chemistry and drug design. However, while recent algorithmic advancements such as qubitization and Tensor Hypercontraction (THC) have significantly reduced the complexity of such calculations, they do not yet achieve computational runtimes short enough to be practical for industrially relevant use cases. In this work, we combine several advances to electronic structure calculation for molecular systems, resulting in a two-orders-of-magnitude speedup of estimated runtimes over prior-art algorithms run on comparable quantum devices. One of these advances is a novel framework for block-invariant symmetry-shifted Tensor Hypercontraction (BLISS-THC), with which we achieve the tightest Hamiltonian factorizations reported to date. We compile our algorithm for an Active Volume (AV) architecture, a technical layout that has recently been proposed for fusion-based photonic quantum hardware. AV compilation contributes towards a lower runtime of our computation by eliminating overheads stemming from connectivity issues in the underlying surface code. We present a detailed benchmark of our approach, focusing primarily on the computationally challenging benchmark molecule P450. Leveraging a number of hardware tradeoffs in interleaving-based photonic FTQC, we estimate runtimes for the electronic structure calculation of P450 as a function of the device footprint.
\end{abstract}

\date{\today}

\maketitle
\tableofcontents
\section{Introduction}
High-accuracy quantum chemical calculations have a large number of industrial applications, ranging from the optimization of reaction rates~\cite{Langer1929, Truhlar1996Reactions} to computer-aided drug design~\cite{Ryde2016Binding, Lam2020AppsCAD, Santagati2024perspective, Baiardi2023QCMolBio}, and battery optimization~\cite{Spotte-Smith2021Battery, Kim2022battery, Delgado2022Battery}. However, because of the exponential scaling of the computational resources for the most accurate methods and the consequent impractical runtimes on classical computers, most quantum chemical calculations are often limited to approximate approaches, such as density functional theory (DFT), which result in less dependable predictions~\cite{JACOBSEN2002_DFTCat, Mardirossian2017_DFT}. Consequently, accurate quantum chemical calculations represent one of the most anticipated practical applications of quantum computers. 

A prime example is the electronic structure calculation of strongly correlated systems~\cite{Cao2019Rev, McArdle2020, Aspuru-Guzik2005, Reiher2017,  Bauer2020, Steudtner2023SEVE}, such as the iron-molybdenum cofactor (FeMoco)~\cite{Reiher2017} or cytochrome P450~\cite{Goings2022P450}, both playing critical roles in biological systems. FeMoco is part of the nitrogenase enzyme, which splits the dinitrogen triple bond, eventually leading to the generation of two ammonia molecules. So, a better understanding of its chemistry could bring new insights to design catalysts for nitrogen fixation. Meanwhile, cytochromes P450 are a group of heme proteins playing a crucial role in the metabolism of drugs. The interplay of drugs with P450 proteins may cause unwanted effects such as drug-drug interactions or accelerated systemic clearance of the active compound from the body. For this reason, they are often considered as anti-targets in computer-aided drug design~\cite{Guengerich2021ToxP450}. While both FeMoco and P450 are relevant to industrial applications and exhibit complex electronic structures, this work primarily focuses on the potential speedups in computing the ground state energy of P450.

In recent years, there has been an increasing endeavor to enhance the existing methods for improving the efficiency of electronic structure calculations in fault-tolerant quantum computing (FTQC). Since the first resource estimates for FeMoco~\cite{Reiher2017}, numerous studies have focused on estimating the upper limits of the computational resources needed, specifically the number of non-Clifford gates and qubits. Thanks to improved quantum algorithms~\cite{Low2016QSP, Childs2012LCU,  Low2019Qubitization, Berry2019Qubitization_published, Gilyen2018QSVT,  Babbush2018Encoding} and better representations of the quantum chemical Hamiltonians~\cite{Lee2020EvenMore, VonBurg2021Catalysis, Rocca2024}, we have witnessed a reduction of several orders of magnitude of the quantum resources required for sampling the eigenspectrum of molecular systems.

From an algorithmic perspective, efforts using quantum phase estimation (QPE) have initially focused on estimating the spectra of electronic structure Hamiltonians $H$, by trotterizing the time evolution operator $U = \exp({-i Ht})$ on a quantum computer~\cite{Kitaev1995, Lloyd1996UQS}. More recent research, however, has shifted towards employing \textit{qubitization}: a version of QPE using a quantum walk operator to encode energies $E$ of $H$ into a spectrum of eigenvalues $\pm  \arccos (E/\lambda)$, where $\lambda$ is a parameter referred to as the \emph{1-norm}. Not to be confused with the induced $\ell_1$ of the Hamiltonian, the 1-norm is an upper bound on the operator norm of $H$. On a technical level, the value of the 1-norm arises in the \emph{block encoding}, the part of the walk operator applying the Hamiltonian to the qubits representing the system~\cite{Childs2012LCU, Low2019Qubitization, Berry2019Qubitization_published}.
Qubitization is regarded as the state-of-the-art for electronic structure calculations, and its computational cost is directly proportional to $\lambda$. To significantly reduce the runtime of the quantum computation, one may use algorithms that encode factorized Hamiltonians with a lower 1-norm. The quantum algorithms with the current best performance (lowest resource cost) rely either on a Double Factorization (DF) approach~\cite{VonBurg2021Catalysis} or make use of Tensor Hypercontraction (THC)~\cite{Lee2020EvenMore}. These methods have shown very encouraging improvements in the expected quantum computing runtime for sampling the spectrum of the electronic structure Hamiltonian, estimated to be approximately $10^2$  hours (or roughly $10^9$ non-Clifford gates) required for both the FeMoco and the P450 systems~\cite{Lee2020EvenMore, Goings2022P450}  using a specific superconducting architecture running surface code. On top of these initial proposals, a number of recent works have made additional improvements to the 1-norm  by combining DF with block-invariant symmetry-shifts (BLISS) of the Hamiltonian~\cite{Rocca2024, Loaiza2023BLISSIzmaylov, Patel2024_BLISS_DF, Berry2024Rapid, Oumarou2024acceleratingquantum,deka2024simultaneously}. 

However, even using the most efficient algorithms available today~\cite{Lee2020EvenMore}, the wallclock runtimes of such calculations for industrially relevant systems such as cytochrome P450~\cite{Goings2022P450} are on the order of days. In~\citet{Santagati2024perspective} it is argued that this is incompatible with the speed required for pharmaceutical industrial workflows, where calculations in the range of seconds or faster would be required. This requirement comes from the fact that, in the majority of cases, the calculation of ensemble properties is of interest. Usually, simulations require the sampling of a large number ($O(10^6)$)~\cite{adcock2006,zwier2010} of single-energy evaluations to compute the properties of thermodynamic ensembles~\cite{Ginex2024QM_in_drugdiscovery}. Currently, severe approximations allow the treatment of relevant-sized systems by quantum chemistry, but for strongly correlated systems such as drugs interacting with the heme center of P450 proteins, these approximations are not applicable. However, adequate descriptions are computationally out of reach of classical hardware. This is why further reductions in computational costs are needed for quantum computers to become relevant to industrial applications. This will likely require combining algorithmic and Hamiltonian decomposition improvements with better hardware and more efficient architectures. Recently, a novel \emph{Active Volume} (AV) architecture has been proposed for fusion-based photonic FTQC, promising a considerable reduction in the computational runtime by exploiting non-local connections between the physical components in the quantum computer~\cite{litinski2022active}. In conventional architectures without Active Volume capabilities,  fault-tolerant operations require the participation of a large number of logical qubits that could otherwise be idle, preventing the compiler from using those qubits for other tasks in parallel. While the performance of AV compilation has been analyzed for RSA factorization~\cite{litinski2022active} and elliptic-curve cryptography~\cite{Litinski2023_Elliptic}, a study for chemistry problems is still lacking.

In this work, we integrate the BLISS technique with THC. Using a P450 system as a benchmark, we demonstrate that BLISS-THC, AV compilation, and some modifications of the block encoding circuit, reduce the computational
runtime of an electronic structure calculation of P450 by at least a factor of $\Mark{233}$, as compared to a THC calculation on an equivalent photonic hardware without AV-capabilities.
A breakdown of the different speedups can be found in Table~\ref{tab:speedups}, and a brief history of previous improvements to electronic structure calculations of P450, leading up to our work, can be found in Table~\ref{tab:factorization}.

To obtain physical runtimes and space requirements, we provide a pragmatic review of the design features of a fault-tolerant quantum computer based on photonic fusions. For instance, in these types of devices, the number of \emph{interleaving modules} (IMs) becomes a key metric to describe the physical size of the quantum computer. For photonic FTQC,  the number of physical qubits, typically referenced in other architectures, has little meaning for the device footprint due to a feature called \emph{interleaving}~\cite{bombin2021interleaving}.  By storing entangled photons in a fiber, interleaving facilitates a tradeoff between the physical runtime and the number of IMs,  thus decoupling the code distance from the physical size of the device.

Our work, therefore, presents physical runtimes as a function of the number of IMs in the quantum computer. For devices with a large amount of interleaving, we obtain the minimum number of interleaving modules required to run the computation in a certain time frame, allowing us to compare our results with prior art~\cite{Goings2022P450}. What is more, we explore additional runtime improvements using algorithmic tradeoffs between the number of qubits assigned to the memory of the computation, and its \emph{workspace}, the group of qubits used for operational tasks. 

The remainder of this paper is organized as follows: in Section \ref{sec:blissthc}, we discuss some of the Hamiltonian factorization techniques referenced. Reviewing THC and BLISS, we finally developed BLISS-THC. In Section~\ref{blockencodingsection}, we introduce the quantum circuit for the block encoding of the new Hamiltonian.  After that, we lay the foundation for hardware and runtime calculations in Section~\ref{sec:hardware}, where we review Active Volume architectures, interleaving,  workspace qubits, and runtime tradeoffs. Drawing from the previous sections, we present our results for the P450 benchmark system in Section~\ref{sec:results}, where we analyze the BLISS-THC Hamiltonian, obtain relative speedups from logical resource counts, provide wallclock runtimes and compute minimal device requirements. Our results feature a number of spacetime tradeoffs, in particular with the code distance and size of the workspace, as devices with smaller code distances / larger workspaces compute faster. This allows us to turn qubits savings due to modifications of the algorithm into speedups.
We conclude with proposals for future research directions in Section~\ref{sec:conclusion}.

\begin{table}[b]
\caption{Contributions to the speedup of the electronic structure calculation of P450, as compared to the THC-based calculation~\cite{Lee2020EvenMore} on a baseline architecture defined in Section~\ref{sec:hardware}. The three steps, 1) Active Volume compilation, 2) the incorporation of BLISS within THC, and 3) improvements to the block encoding circuit, are done in sequence while adding logical qubits saved in the memory to the workspace. Speedups reported reflect the runtime improvements with respect to the previous step. Multiplying the individual improvements hence results in the total speedup, up to rounding errors: all numbers in this table are rounded down, but the total speedup is computed with the exact estimates. }\vspace{.5cm}
\begin{tabular}{l@{\hskip 10pt}r}
\toprule
Method & Speedup factor \\
\hline
AV compilation & $25.18\times$ \\
THC $\mapsto$ BLISS-THC  &  $8.23\times$\\
Circuit  improvements & $1.12\times$\\
\hline
Total: & $233.96\times $\\
\bottomrule
\end{tabular}
\label{tab:speedups}
\end{table}

\begin{table*}[tb]
\centering
\caption{Logical qubit requirements for the device memory, Toffoli counts, Active Volume count (where applicable), as well as the 1-norm for electronic structure calculations of P450 in a (63e, 58o) active space configuration with respect to Double Factorization (DF), Symmetry-Compressed Double Factorization (SCDF), Tensor Hypercontraction (THC) and BLISS-THC, the product of this work. For our efforts, we have obtained tighter estimates for the costs of the THC circuit with respect to the parameters of prior art. Note that BLISS-THC approaches $69.3\,\mathrm{E_h}$, the theoretical 1-norm limit for P450~\cite{Cortes2024}, to within a factor of 2.} \vspace{0.5cm}
\begin{tabular}{l@{\hskip 10pt}r@{\hskip 10pt}l@{\hskip 10pt}c@{\hskip 10pt}r}
\toprule
\multirow{2}{*}{Factorization} & \multicolumn{4}{c}{P450 (63e, 58o)} \\
& Memory qubits & Toffolis & Active Volume & {$\lambda$ ($\mathrm{E_h}$)} \\
\hline
DF~\cite{Rocca2024} & 4,922 & 1.9$\times 10^{10}$ & --- & 472.2  \\
SCDF~\cite{Rocca2024} & 1,706 & 4.8$\times 10^{9}$ & --- & 111.3 \\
THC~\cite{Goings2022P450} &1,434 & 7.8$\times 10^{9}$   & --- & 388.9 \\
THC [re-estimated for this work] & 1,357 &  7.8$\times 10^9$ & 1.6$\times 10^{12}$  & 388.9\\ 
BLISS-THC [this work] & 999 & 1.7$\times 10^{9}$  & 2.3$\times 10^{11}$ & 130.9 \\
\bottomrule
\end{tabular}
\label{tab:factorization}
\end{table*}

\section{BLISS-THC}
\label{sec:blissthc}
\subsection{Theoretical Overview}
Within the standard framework of second quantization, we consider the electronic structure Hamiltonian written in chemist notation,
\begin{align}
	H &= \sum_{\substack{pq \\\sigma}}\left( h_{pq} -\frac{1}{2}\sum_r g_{prrq}\right)  a^\dagger_{p\sigma} a^{\phantom{\dagger}}_{q\sigma} \nonumber\\
	&\quad\quad+ \frac{1}{2}\sum_{\substack{pqrs \\\sigma\tau}} g_{pqrs} a^\dagger_{p\sigma}a^{\phantom{\dagger}}_{q\sigma} a^\dagger_{r\tau} a^{\phantom{\dagger}}_{s\tau} ,
    \label{electronic_hamiltonian}
\end{align}
where $\{\sigma,\tau\}$ are indices for spin configurations $\uparrow, \downarrow$ and $\{p,q,r,s\}$ are indices labeling spatial orbitals. The terms $h_{pq}$ and $g_{pqrs}$ are the conventional 1-body and 2-body integrals defined with real molecular spatial orbitals $\phi_p(\mathbf{r})$,
\begin{align}
    h_{pq} &= \int \mathrm{d}\mathbf{r}_1 \phi_p(\mathbf{r}_1)\left( -\frac{1}{2}\nabla^2 - \sum_I \frac{Z_I}{r_I} \right)\phi_q(\mathbf{r}_1) , \\
    g_{pqrs} &= \iint \mathrm{d}\mathbf{r}_1 \mathrm{d}\mathbf{r}_2 \frac{\phi_p(\mathbf{r}_1)\phi_q(\mathbf{r}_1)\phi_r(\mathbf{r}_2)\phi_s(\mathbf{r}_2)}{|\mathbf{r}_1-\mathbf{r}_2|}.
\end{align}
Given the symmetry-shifting techniques to be introduced in the following sections, we also emphasize the symmetries of the electronic structure Hamiltonian, including total particle number $\hat{N}$, spin projection $\hat{S}_z$, and total spin $\hat{S}^2$,
\begin{align}
    \hat{N}   &= \sum_p ( a^\dagger_{p\uparrow} a^{\phantom{\dagger}}_{p\uparrow} + a^\dagger_{p\downarrow} a^{\phantom{\dagger}}_{p\downarrow} ) \, , \\
    \hat{S}_z &= \tfrac{1}{2}\sum_p ( a^\dagger_{p\uparrow} a^{\phantom{\dagger}}_{p\uparrow} - a^\dagger_{p\downarrow} a^{\phantom{\dagger}}_{p\downarrow} ) \, , \\
    \hat{S}^2 &= \hat{S}_+\hat{S}_- + \hat{S}_z(\hat{S}_z - 1) \, ,
\end{align}
where $\hat{S}_+ = \sum_p a^\dagger_{p\uparrow} a^{\phantom{\dagger}}_{p\downarrow}$ and $\hat{S}_- = (\hat{S}_+)^\dagger$. While other symmetries may appear in specific problem instances, the symmetries mentioned above are always applicable.

\subsection{Tensor Hypercontraction}

This work uses the Tensor Hypercontraction (THC) factorization technique to decompose the four-index tensor, $g_{pqrs}$~\cite{parrish2012tensor,hohenstein2012tensor}. THC is currently known to provide the best asymptotic scaling for ground state energy estimation based on qubitized quantum phase estimation with $\tilde{O}(N)$ logical qubits and $\tilde{O}(N\lambda_\mathrm{THC}/\epsilon)$  Toffoli gates. In this framework, the four-index tensor, $g_{pqrs}$, is factorized using two-index tensors $\zeta_{\mu\nu}$ and $\chi^\mu_p $,
\begin{align}
    g_{pqrs} = \sum_{\mu,\nu=0}^{M-1}  \zeta_{\mu\nu} \,\chi^\mu_p \chi^\mu_q  \chi_r^\nu \chi_s^\nu,
    \label{THC_expansion}
\end{align}
where $\zeta_{\mu\nu}$ is a real symmetric matrix and $M$ is the factorization rank. The tensor $\chi^{\mu}_p$ is also assumed to be normalized such that $\sum_p \chi^\mu_p \chi^\mu_p = 1$. Combining Eqs.~\eqref{THC_expansion} and~\eqref{electronic_hamiltonian} leads to the following expanded form of the Hamiltonian,
\begin{widetext}
\begin{align}
    H\;  =  &\; \sum_{\substack{pq \\\sigma}} \left( h_{pq} -\frac{1}{2}\sum_r g_{prrq}\right)a^\dagger_{p\sigma} a^{\phantom{\dagger}}_{q\sigma}  + \; \frac{1}{2}\sum_{\substack{\mu\nu \\ \sigma\tau}} \zeta_{\mu\nu} \left(\sum_p \chi^\mu_p a^\dagger_{p\sigma}\right)\left(\sum_q \chi^\mu_q a^{\phantom{\dagger}}_{q\sigma} \right)\left(\sum_r \chi^\nu_r a^\dagger_{r\tau}  \right)\left(\sum_s \chi^\nu_s a^{\phantom{\dagger}}_{s\tau}\right).
    \label{THC_Hamiltonian}
\end{align}
\end{widetext}
The second term can be simplified considerably by defining a linear transforms $\sum_p \chi^\mu_p a_{p\sigma} = U_\mu^\dagger a_{0\sigma} U_\mu$, where the unitaries $U_\mu$ are products of Givens rotation operators that we cover later. The Jordan-Wigner transformation then gives rise to  the Hamiltonian~\cite{Lee2020EvenMore}:
\begin{align}
	H &= -\frac{1}{2}\sum_{\substack{k,\sigma}} t_k\, V_k^\dagger Z_{0,\sigma} V^{\phantom{\dagger}}_k\, \nonumber\\
	&\quad\quad + \frac{1}{8}\sum_{\substack{\mu\nu \\\sigma\tau}} \zeta_{\mu\nu} U^\dagger_\mu Z_{0,\sigma} U_\mu^{\phantom{\dagger}} U_\nu^\dagger Z_{0,\tau} U_\nu^{\phantom{\dagger}},
    \label{JW_THC_Hamiltonian}
\end{align}
where $t_k$ are the eigenvalues of the one-body matrix $T_{pq} =  h_{pq} -\frac{1}{2}\sum_r g_{prrq} + \sum_r g_{pqrr}$, while $Z_{0,\sigma}$ is a Pauli $Z$ operator with respect to the topmost qubit in the register associated with the spin index $\sigma$, and $V_k$ is the orbital rotation operator associated with $t_k$, defined similarly to $U_\mu$. A block encoding capable of implementing the basis-rotated $Z$ operators, would yield the 1-norm,
\begin{align}
    \lambda = \sum_k \left|t_k\right| + \frac{1}{2}\sum_{\mu\nu}\left|\zeta_{\mu\nu}\right|.
\end{align}
While this 1-norm expression is correct for the quantum circuit implementation proposed in~\cite{Lee2020EvenMore}, it is possible to show that a small modification of the circuit leads to the slightly smaller 1-norm, 
\begin{equation}
    \lambda_\circ = \sum_k \left|t_k\right| + \frac{1}{2}\sum_{\mu\nu}\left|\zeta_{\mu\nu}\right| - \frac{1}{4}\sum_\mu \left|\zeta_{\mu\mu}\right|. 
\end{equation}
This modification involves a special treatment of the diagonal terms  $\tilde{\zeta}_{\nu\nu}$ in Eq.~\eqref{JW_THC_Hamiltonian}, ensuring that we do not encode  constant terms in our Hamiltonian: for identical spins $\sigma = \tau $ the terms $U_\nu^\dagger Z_{0,\sigma}  U_\nu^{\phantom{\dagger}} U_\nu^\dagger Z_{0,\tau}  U_\nu^{\phantom{\dagger}}$ are equal to $1$. Constant contributions to the Hamiltonian should be avoided, as they would needlessly increase the 1-norm. Our correction leads to an updated Hamiltonian,
\begin{align}
	H_{\circ} &= -\frac{1}{2}\sum_{\substack{k,\sigma}} t_k\, V_k^\dagger Z_{0,\sigma} V^{\phantom{\dagger}}_k\, \nonumber\\
	&\quad\quad + \frac{1}{8}\sum_{\substack{\mu\nu \\\sigma\tau}} \zeta_{\mu\nu} U^\dagger_\mu Z_{0,\sigma} U_\mu^{\phantom{\dagger}} U_\nu^\dagger Z_{0,\tau} U_\nu^{\phantom{\dagger}} -\frac{1}{4}\sum_{\mu} \zeta_{\mu\mu}.
    \label{JW_THC_Hamiltonian_new}
\end{align}

\subsection{Block-invariant symmetry-shift (BLISS) framework}

The block-invariant symmetry-shift (BLISS)~\cite{Loaiza2023BLISSIzmaylov, Patel2024_BLISS_DF} framework defines a block-invariant Hamiltonian, $H_\mathrm{BI}$, that shifts the original Hamiltonian by a well-behaved symmetry function, $f(\hat{N},\hat{S}^2,\hat{S}_z)$, leaving the eigenvalue spectrum of the desired symmetry sector unchanged apart from shifts with a constant. This strategy changes the eigenvalues of other symmetry sectors in order to minimize the 1-norm $\lambda$. Explicitly, the block-invariant Hamiltonian may be defined as, 
\begin{align}
    H_\mathrm{BI} &= H - \alpha_1 \hat{N} - \frac{\alpha_2}{2}\hat{N}^2  - \frac{1}{2}\hat{B}( \hat{N} - \eta )
    \label{BI_Hamiltonian}
\end{align}
where the second and third terms are equal to 1-body and 2-body symmetry-shift operators, based on the total particle number operator $\hat{N}$, obeying the eigenvalue relation, $\hat{N}\ket{\psi_\eta} = \eta\ket{\psi_\eta}$, where $\eta$ is the eigenvalue of the particle number operator, equal to the total number of spin-up and spin-down electrons, and $\ket{\psi_\eta}$ is any electronic wavefunction defined in the $\eta$-electron symmetry sector. The fourth term in Eq.~\eqref{BI_Hamiltonian} is the BLISS operator equal to zero in the desired particle number sector and non-zero everywhere else, where $\hat{B}$ is a Hermitian 1-body operator defined as
\begin{equation}
    \hat{B} = \sum_{\substack{pq \\\sigma}} \beta_{pq} a^\dagger_{p\sigma} a^{\phantom{\dagger}}_{q\sigma}.
\end{equation}
In total, the coefficients $\alpha_1, \alpha_2$ and $\beta_{pq}$ encode a symmetry-based gauge invariance, which helps to reduce the 1-norm compared to what is theoretically achievable with the conventional Hamiltonian~\cite{Loaiza_2023_LCU}. The BLISS method ultimately exploits the fact that the conventional Hamiltonian $H$ inherently includes redundant information, which is only necessary when considering the full Fock space with all its possible symmetry sectors. Within the context of this quantum circuit, the input and output wavefunctions will generally belong to a single, well-defined symmetry sector, implying that the compiled Hamiltonian must only be equal to the original Hamiltonian within the very same sector. We conclude this section by explicitly writing the block-invariant Hamiltonian as, 
\begin{align}
    H_\mathrm{BI}    &= \sum_{\substack{pq \\ \sigma}} h^\mathrm{(BI)}_{pq} a_{p\sigma}^\dagger a^{\phantom{\dagger}}_{q\sigma} + \tfrac{1}{2}\sum_{\substack{pqrs \\ \sigma\tau}} g^\mathrm{(BI)}_{pqrs} a^\dagger_{p\sigma} a^{\phantom{\dagger}}_{q\sigma} a^\dagger_{r\tau} a^{\phantom{\dagger}}_{s\tau},
\end{align}
where we define the renormalized block-invariant integrals as
\begin{align}
    h^\mathrm{(BI)}_{pq} &=  h_{pq} -\frac{1}{2}\sum_r g_{prrq} - \alpha_1\delta_{pq} + \tfrac{1}{2}\beta_{pq}\eta  \\
    g^\mathrm{(BI)}_{pqrs} &= g_{pqrs} - \alpha_2\delta_{pq}\delta_{rs} - \tfrac{1}{2}(\beta_{pq}\delta_{rs}+\delta_{pq}\beta_{rs}).
\end{align}
In the following section, we build off this work but perform the analysis in the Majorana representation, which is required to properly implement the THC framework for the block encoding.

\subsection{Block-invariant symmetry-shifted Tensor Hypercontraction}
While recent work \cite{Cortes2024} has shown that previous implementations of THC are within a factor of 2 to 3 of the conventional spectral bound, which is symmetry-agnostic, they are still within a factor of 4 to 5 away from the bound in the $S=5/2$ symmetry sector of P450.
We improve upon those results by proposing a block-invariant symmetry-shifted Tensor Hypercontraction: BLISS-THC. 
To provide a unified framework that works both with the standard THC representation as well as BLISS-THC, we present the results of this section using the Majorana representation of the Hamiltonian, where $a_{p\sigma}=(\hat{\gamma}_{p\sigma,0} + i \hat{\gamma}_{p\sigma,1})/2$ and $a^\dagger_{p\sigma}=(\hat{\gamma}_{p\sigma,0} - i \hat{\gamma}_{p\sigma,1})/2$. This representation is preferred due to the properties of Majorana operators, which are Hermitian and self-inverse, and have a clear one-to-one mapping with Pauli strings after performing the Jordan-Wigner transformation. Our main result is the block-invariant Majorana-based electronic structure Hamiltonian,
\begin{align}
	H_\mathrm{BI}  &= \frac{i}{2} \sum_{\substack{pq\\\sigma}}\kappa^\mathrm{(BI)}_{pq}\hat{\gamma}_{p\sigma,0}\hat{\gamma}_{q\sigma,1} \nonumber\\
	&\quad\quad- \frac{1}{8}\sum_{\substack{pqrs\\\sigma\tau}} g^\mathrm{(BI)}_{pqrs} \hat{\gamma}_{p\sigma,0}\hat{\gamma}_{q\sigma,1} \hat{\gamma}_{r\tau,0}\hat{\gamma}_{s\tau,1},
\end{align}
where
\begin{align}
    \kappa^\mathrm{(BI)}_{pq} &= h_{pq}-\frac{1}{2}\sum_r g_{prrq} + \sum_r g^\mathrm{(BI)}_{pqrr} \nonumber\\
	&\quad\quad- \alpha_1\delta_{pq} + \tfrac{1}{2}\beta_{pq}\eta,      \\
    g^\mathrm{(BI)}_{pqrs} &= g_{pqrs} - \alpha_2\delta_{pq}\delta_{rs} - \tfrac{1}{2}(\beta_{pq}\delta_{rs}+\delta_{pq}\beta_{rs}).
\end{align}
These coefficients correctly encode the symmetry-shift and BLISS operations, renormalizing the coefficients that are typically found in the Majorana representation~\cite{VonBurg2021Catalysis}. In the BLISS-THC framework, the block-invariant four-index tensor, $g^\mathrm{(BI)}_{pqrs}$, is factorized as,
\begin{align}
    g^\mathrm{(BI)}_{pqrs} = \sum_{\mu\nu}^M  \tilde{\zeta}_{\mu\nu} \,\chi^\mu_p \chi^\mu_q  \chi_r^\nu \chi_s^\nu\, ,
\end{align}
which, combined with the elimination of constant terms, gives rise to the block-invariant THC Hamiltonian,
\begin{align}
    \widetilde{H} &= -\frac{1}{2} \sum_{\substack{k \\\sigma}} \tilde{t}_k V^\dagger_k Z_{0,\sigma} V_k\nonumber\\
	&\quad\quad    + \frac{1}{8}\sum_{\substack{\mu\nu\\\sigma\tau}} \tilde{\zeta}_{\mu\nu} U^\dagger_\mu Z_{0,\sigma} U_\mu U^\dagger_\nu Z_{0,\tau} U_\nu - \frac{1}{4}\sum_\mu \tilde{\zeta}_{\mu\mu}, 
    \label{eq:biham}
\end{align}
where $\tilde{t}_k$ are the eigenvalues of the one-body matrix $\kappa^\mathrm{(BI)}_{pq}$. Since the proposed implementation, as defined above in Eq.~\eqref{eq:biham}, does not alter the final form of the Hamiltonian when compared to Eq.~\eqref{JW_THC_Hamiltonian}, we find the final 1-norm expression of the BLISS-THC Hamiltonian is similarly given by,
\begin{equation}
\label{eq:blissthclambda}
    \tilde{\lambda}_\mathrm{THC} = \sum_k |\tilde{t}_k| + \frac{1}{2}\sum_{\mu\nu}|\tilde{\zeta}_{\mu\nu}| - \frac{1}{4}\sum_\mu |\tilde{\zeta}_{\mu\mu}|. 
\end{equation}

\section{Quantum circuits}
\label{blockencodingsection}
\begin{figure*}[tb]
\begin{tikzpicture}
    \node at (-1,0){\includegraphics[width=.47\linewidth]{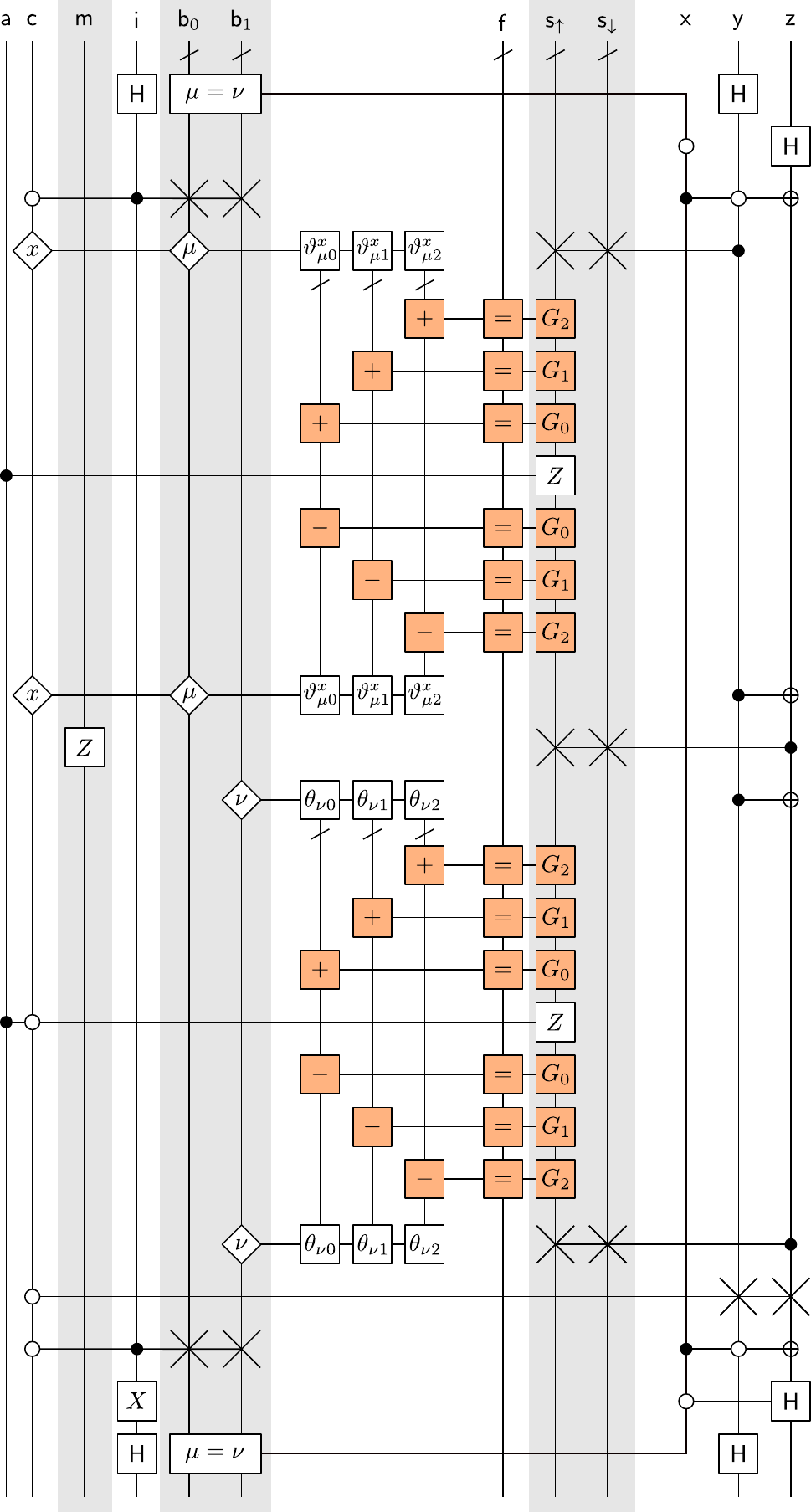}};
    \node at (8, .6) {\begin{tabular}{rc p{0.3\linewidth}}
    $\mathsf{a}$  &\dots& Single qubit, flagging success in the $\mathsf{Prepare}$ circuit with $|1\rangle$.   \\
    $\mathsf{b}_0, \mathsf{b}_1$ &\dots& Registers with  $\lceil\log(M+1)\rceil$ qubits each, encoding the numbers $\mu$ and $\nu$. \\
    $\mathsf{c}$ &\dots& Single qubit flagging the case where $\nu = M$, which indicates that $\phi_{\mu k}$, the angles of the one-body terms have to be loaded, rather than two-body angles $\theta_{\mu k}$. The qubit also disables the second operator $Z_{0,\tau}$ when set.\\
    $\mathsf{f}$ &\dots& Register holding a $(\beth+1)$-qubit phase gradient state $\propto\sum_{k} \exp(i\pi k/2^{\beth}) |k\rangle$. \\
    $\mathsf{i}$ &\dots& Single qubit that is utilized to flag conditions under which the registers $\mathsf{b}_0$ and $\mathsf{b}_1$ are swapped, a procedure that is necessary as $|\Lambda\rangle$ has the restriction $\mu\leq \nu$ on its indices $|\mu\rangle|\nu\rangle$, but Eq.~\eqref{eq:biham} requires all configurations of $\mu$ and $\nu$, including the cases where $\mu>\nu$. This qubit plays a pivotal role in rendering the $\mathsf{Select}$ circuit self-inverse.  \\
    $\mathsf{m}$ &\dots& Single qubit that is set to $|1\rangle$ in the subspaces where the coefficients $\tilde{\zeta}_{\mu\nu}$ are negative. This qubit can be made redundant when the sign is implemented in the $\mathsf{Prepare}$ circuit, as shown in Fig.~4 of~\cite{Lee2020EvenMore}.\\
    $\mathsf{s}_{\uparrow}, \mathsf{s}_\downarrow$ &\dots& Two $N/2$-qubit registers representing all the spin-up/-down orbitals of the system, respectively. \\
    $\mathsf{x}$ &\dots& Qubit flagging the case where $\mu=\nu$, utilized to exclude constant terms in the Hamiltonian in the case where $\sigma=\tau$.\\
    $\mathsf{y}, \mathsf{z}$ &\dots& Single qubits that swap the registers $\mathsf{s}_{\uparrow}$ and $\mathsf{s}_{\downarrow}$ when set, accounting for the spin variables $\sigma$ and $\tau$ in Eq.~\eqref{eq:biham}.
 \end{tabular}};
 \node at (-1,8.1) {$\boldsymbol{(a)}$ $\mathsf{Select}$ circuit};
 \node at (8,8.1) {$\boldsymbol{(b)}$ Quantum registers};
\end{tikzpicture}
    \caption{$\boldsymbol{(a)}$ A small $\mathsf{Select}$ circuit of BLISS-THC for $N=8$. The circuit is to be read from top to bottom, where the $Z$ gate refers to a Pauli $Z$ operator on the topmost qubit of $\mathsf{s}_\uparrow$, $\mathsf{H}$ are Hadamard gates, and the gates labeled `$\mu=\nu$' compare the values stored in both registers and flip the target qubit if the integers match. The gates with the $\diamond$-shaped connectors are dataloaders, loading the angles $\theta_{jk}$ and $\vartheta^x_{jk}$ into freshly allocated qubits, where we adopt the elbow formalism of \citet{Babbush2018Encoding}. Here the angles $\vartheta^x_{jk}$ are the product of a combined register such that the angles $\vartheta^0_{jk}=\theta_{jk}$ are associated with 2-body operators $U_j$ and $\vartheta^1_{jk}=\phi_{jk}$ are associated with 1-body operators, following Eqs.~\eqref{eq:linops}. These angles serve as inputs for the programmable  Givens rotation circuits, highlighted in orange and depicted in detail within Figure~\ref{fig:givens}$(b)$, involving a phase gradient state in register $\mathsf{f}$. Note that the $X$ gate applied to qubit $\mathsf{i}$ is important to make $\mathsf{Select}$ self-inverse i{.}e{.} $(\mathsf{Select})^2=1$, an important prerequisite for qubitization. $\boldsymbol{(b)}$ Quantum registers in the $\mathsf{Select}$ circuit and the auxiliary state $|\Lambda\rangle$ in Eq.~\eqref{aux_state}, not accounting for garbage registers of Figure~\ref{fig:prepare}.} 
    \label{fig:select}
\end{figure*}
In this section, we consider the Hamiltonian block encoding of BLISS-THC, a unitary quantum circuit that acts as the factorized Hamiltonian in a certain subspace of the quantum system. Since BLISS-THC and THC produce a Hamiltonian of the same form, we propose a quantum circuit quite similar to the one in \cite{Lee2020EvenMore}, but with a few modifications that prove to be beneficial for its complexity. Since the electronic structure calculation consists almost entirely of calls to the walk operator, even small modifications of the block encoding can have an impact on the overall runtime.

In the following, we provide a pedagogical explanation of the circuit, highlight our modifications to its construction, and refer the reader to the literature for further details. The block encoding is a sequence
$$
\mathsf{Prepare}^{\dagger\phantom{\dagger}} \mathsf{Select}^{\phantom{\dagger}} \mathsf{Prepare}^{\phantom{\dagger}} 
$$
of the subroutines $\mathsf{Select}$, depicted in Figure~\ref{fig:select}, and $\mathsf{Prepare}$, depicted in Figure~\ref{fig:prepare},  as well as the uncomputation of the latter. Let us start with the $\mathsf{Prepare}$ routine in Section~\ref{preparesection}. After learning about the routine and the qubit registers featured in this algorithm, we move on to $\mathsf{Select}$, and first focus on the part of the circuit implementing the 1- and 2-body operators in Section~\ref{givensrotsection}:  circuits for the so-called \emph{Givens rotations} are highlighted orange within Figure~\ref{fig:select}.  In Section~\ref{selectsection} we consider the full $\mathsf{Select}$ circuit as well as some of its variations that can facilitate tradeoffs between time and space complexity.

\subsection{Prepare}
\label{preparesection}
For the block encoding of BLISS-THC, we use the $\mathsf{Prepare}$ routine of Lee~\emph{et al.}\ \cite{Lee2020EvenMore} to build an auxiliary state $|\Lambda\rangle$, as well as a $\mathsf{Select}$ routine entangling the auxiliary register with the molecular system qubits, such that
\begin{align}
\label{eq:blocking}
\langle\Lambda|\mathsf{Select}|\Lambda\rangle  \; = \; \frac{\widetilde{H}}{ \tilde{\lambda}_\mathrm{THC}} \, .
\end{align}
In this work, we represent the electron system with the qubit registers $\mathsf{s}_\uparrow$ and $\mathsf{s}_\downarrow$, associated with spin-up and -down orbitals, respectively. A list of all qubits and registers can be found in Figure~\ref{fig:select}$(b)$. The auxiliary state takes the form
\begin{widetext}
\begin{align}
   |\Lambda\rangle \; = \;   &\left( \sqrt{p} |1\rangle_{\mathsf{a}} \sum_{\nu=0}^{M} \sum_{\mu=0}^{\nu} \sqrt{\left|\widehat{\zeta}_{\mu\nu}\right|} \; |\mu\rangle_{\mathsf{b}_0} |\nu\rangle_{\mathsf{b}_1} |\delta_{M,\nu}\rangle_{\mathsf{c}} |\mathrm{sign}(\widehat{\zeta}_{\mu\nu})\rangle_{\mathsf{m}}    \;\;  + \;\;  \sqrt{1-p} |0\rangle_{\mathsf{a}} \dots\right)  |0\rangle_{\mathsf{i}}|0\rangle_{\mathsf{x}}|0\rangle_{\mathsf{y}}|0\rangle_{\mathsf{z}}
   \label{aux_state}
\end{align}
\end{widetext}
 where 
\begin{itemize}
    \item the coefficients $\widehat{\zeta}_{\mu\nu}$ are normalized  version of the 1- and 2-body operators, respectively;
    \item $p$ is understood as a success parameter designed to be close to 1: if qubit $\mathsf{a}$ were to be measured, $p$ would correspond to the probability of projecting into the subspace where the block encoding is  prepared exactly;
    \item the function $x\mapsto \mathrm{sign}(x)$ outputs $0$ unless  $x$ is negative in which case it outputs $1$;
    \item $|\mu\rangle$ and $|\nu\rangle$ denote  computational basis states representing the integers $\mu$ and $\nu$ in binary representation;
    \item $\delta_{M,\nu}$ is a Kronecker delta.\end{itemize}
To achieve the block-encoding property of Eq.~\eqref{eq:blocking}, the coefficients $\widehat{\zeta}_{\mu\nu}$, are matched to the parameters of the Hamiltonian in Eq.~\eqref{eq:biham} as 
\begin{align}
\label{eq:coeffs}
    \widehat{\zeta}_{\mu\nu} = \frac{1}{\tilde{\lambda}_\mathrm{THC}} \times \left\lbrace \begin{array}{lcl}
        \vphantom{\frac{\sum_k}{\sum_k}}
	    -\tilde{t}_{\mu} && \text{for}\;\,\nu = M \;\, \\&&\text{and} \;\,    \mu \in [0, N/2-1] \\ 
        \vphantom{\frac{\sum_k}{\sum_k}}
	    \tilde{\zeta}_{\mu\nu} && \text{for}\;\,\mu, \nu \in [0,M) \;\,\\&&\text{and} \;\, \nu> \mu \\
        \vphantom{\frac{\sum_k}{\sum_k}}
	    \tilde{\zeta}_{\nu\nu} / 4  &&  \text{for}\;\,\nu \in [0,M) \;\,\\&& \text{and} \;\, \mu=\nu  \\ 
        \vphantom{\frac{\sum_k}{\sum_k}}
	    0 &&  \text{else.} 
    \end{array} 
    \right.
\end{align}
The interested reader may find a detailed proof of this mapping in Appendix \ref{sec:blencoding}.
We again encounter a special treatment of the diagonal terms  $\tilde{\zeta}_{\nu\nu}$ in Eq.~\eqref{eq:coeffs}, which is due to our version of $\mathsf{Select}$ excluding the constant terms, that have been subtracted off in Eq.~\eqref{eq:biham}.

Note that to achieve the  1-norm  of Eq.~\eqref{eq:blissthclambda}, we need to make sure the success amplitude $p$ is close to unity: $p\mapsto 1$. Later, spurious contributions to the energy from the inexact subspace of $|\Lambda\rangle$ will be eliminated by controlling the $\mathsf{Select}$ circuit on qubit $\mathsf{a}$.

\begin{figure}
    \centering
    \includegraphics[width=.7\linewidth]{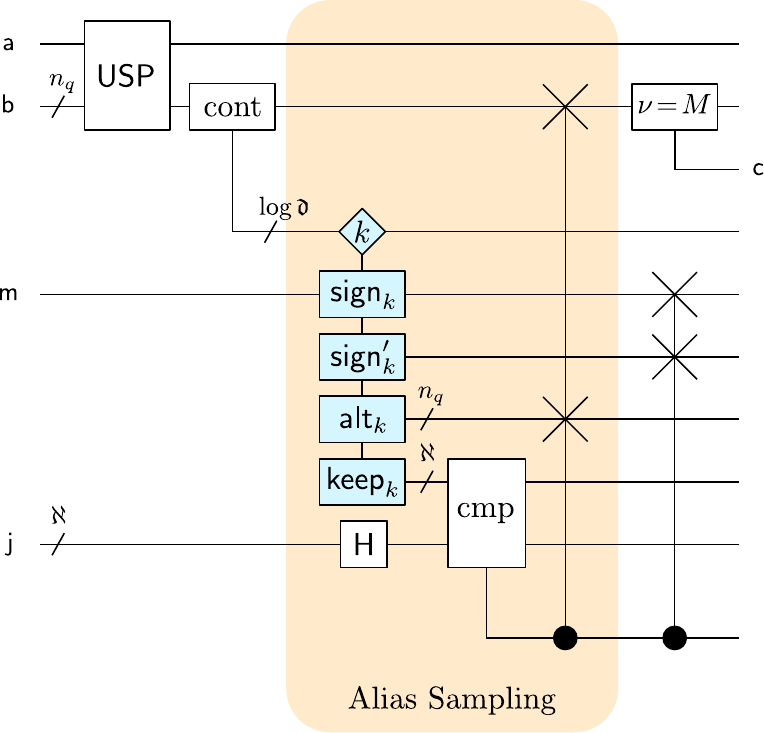}
    \caption{The $\mathsf{Prepare}$ circuit of \cite{Lee2020EvenMore}, where the registers $\mathsf{b}_0$ and $\mathsf{b}_1$ are combined into a register $\mathsf{b}$ of $n_q$ qubits. The register is then fed into the subroutine $\mathsf{USP}$, which prepares a uniform superposition $\propto\sum_{\mu\nu} |\mu\rangle|\nu\rangle$ over all $\mathfrak{d}=M(M+1)/2 + N/2$ values of $\mu\leq\nu$ where $\widehat{\zeta}_{\mu\nu}\neq 0$. This uniform state preparation uses amplitude amplification, and success is therefore flagged in qubit $\mathsf{a}$. This superposition state is a necessary input for the Alias Sampling block (orange) of $\mathsf{Prepare}$, but since the basis states of the superposition do not represent numbers from a continuous range, parsing them would be hard for the routine. An arithmetic contiguizer ($\mathrm{cont}$) therefore entangles the basis states $|\mu\rangle|\nu\rangle$ of $\mathsf{b}$ with basis states $|k(\mu,\nu)\rangle$ of a temporary register, $|\mu\rangle|\nu\rangle|0\rangle\mapsto|\mu\rangle|\nu\rangle|k(\mu,\nu)\rangle$, such that the numbers $k(\mu,\nu)$ are covering all integers from $0$ to $\mathfrak{d}-1$ within the superposition. Inside the Alias Sampling routine, a dataloader (blue) loads $\aleph$-bit $\mathsf{keep}_k$ values and alternative indices $\mathsf{alt}_k = (\iota,\kappa)$ to the index pair $(\mu,\nu)$ of a contiguous value $k(\mu,\nu)$. The routine also loads $\mathsf{sign}_k$, the sign of the corresponding coefficient $\widehat{\zeta}_{\mu\nu}$  as well as the sign $\mathsf{sign}^\prime_k$ of the coefficient $\widehat{\zeta}_{\iota\kappa}$ with respect to the alternative indices. A comparator $(\mathrm{cmp})$ flags the cases where the values $|m\rangle$ of a register $\mathsf{j}$ are larger than the $\mathsf{keep}_k$ values. In this case, the alternative indices are swapped into $\mathsf{b}$. The values of qubit $\mathsf{c}$ are computed after the Alias Sampling, using a controlled swap and the comparator `$\nu=M$', Here $\mathsf{H}$ denotes the application of Hadamard gates to every qubit in a register. The register $\mathsf{j}$ and temporary qubits make up the garbage register after the circuit.}
    \label{fig:prepare}
\end{figure}

To encode $|\Lambda\rangle$, we follow the procedure of Lee~\emph{et al.}\ and utilize a modified \emph{Alias Sampling} routine \cite{Babbush2018Linear} to prepare the state
$$
\sum_\nu\sum_\mu \sqrt{\left| \widehat{\zeta}_{\mu\nu}\right|} \;  |\mu\rangle_{\mathsf{b}_0} |\nu\rangle_{\mathsf{b}_1}\; |\text{trash}_{\mu\nu}\rangle_{\mathsf{g}} 
$$
setting the values of qubits $\mathsf{c}$ and $\mathsf{m}$ in the process. In the resulting version of $|\Lambda\rangle$, the registers $\mathsf{b}_0$ and $\mathsf{b}_1$ are entangled with some states $|\text{trash}_{\mu\nu}\rangle$ in a garbage register $\mathsf{g}$, but this is an acceptable overhead and has no consequence for the calculation. Before entering  the Alias Sampling routine, both registers $\mathsf{b}_0$ and $\mathsf{b}_1$ must be prepared in a uniform superposition of all configurations $|\mu\rangle |\nu\rangle$ with $\nu \leq M$ and $\mu \leq \nu$. Starting from a superposition created by applying Hadamard gates to all qubits in $\mathsf{b}_0$ and $\mathsf{b}_1$, which would certainly include states with $\nu > M$ and $\mu > \nu$, we can obtain the desired superposition after at most two rounds of amplitude amplification featuring comparator circuits and partial reflections, where  $\sqrt{1-p}$, the amplitude of the residual state with $\nu > M$ and $\mu > \nu$, can be made exponentially small by increasing the bit precision of the synthesized rotations. The $\mathsf{Prepare}$ circuit of~\cite{Lee2020EvenMore} is depicted in Figure~\ref{fig:prepare}.

Note that the dataloader, highlighted blue in Figure~\ref{fig:prepare}, is generally made optimal with respect to its magic state cost,  by turning the QROM into a QROAM using the techniques developed in~\cite{Low2018TradingT}. With $\aleph$ qubits assigned to register $\mathsf{j}$, which is later absorbed into the garbage register $\mathsf{g}$, $\mathsf{Prepare}$ can approximate the coefficients $\widehat{\zeta}_{\mu\nu}$ within the accuracy
\begin{align}
    \frac{2^{-\aleph + 1}}{M(M+1) + N}\, .
\end{align}

\begin{figure}[tb]
    \centering
    \includegraphics[width=\linewidth]{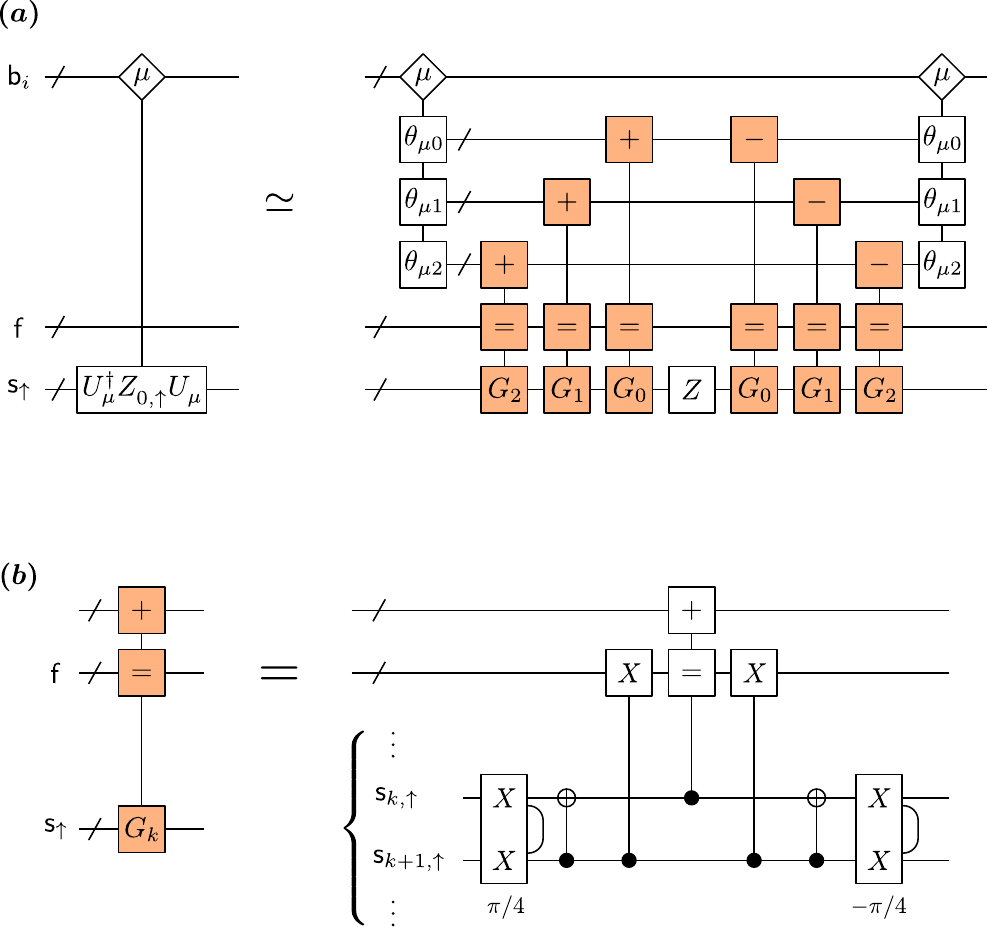}
    \caption{$\boldsymbol{(a)}$~The $\mathsf{subselect}_i$ circuit, depicted on the right-hand side, corresponds to a multiplexed application of $U_\mu^\dagger Z_{0,\uparrow} U_\mu^{\phantom{\dagger}}$ (on the left). The circuit loads the angles $\theta_{\mu,k}$ into temporary registers and employs programmable Givens rotation circuits (orange), where both $Z$ and $Z_{0,\uparrow}$ refer to Pauli $Z$ operators on the topmost qubit of the register $\mathsf{s}_\uparrow$. In this minimal example, $\mathsf{s}_\uparrow$ has only $N/2=4$ qubits. $\boldsymbol{(b)}$~A programmable Givens rotation subcircuit. Here, the lower register, $\mathsf{s}_\uparrow$ fans out into singular qubits $\mathsf{s}_{0,\uparrow}, \mathsf{s}_{1,\uparrow},  ..., \mathsf{s}_{N/2-1,\uparrow}$ of which this circuit uses exactly two subsequent qubits, determined by the index $k$ of $G_k$.  The circuit implements the product of rotations $\exp(-i\pi\theta X_{\mathsf{s}_{k+1,\uparrow}}Y_{\mathsf{s}_{k, \uparrow}})$ and $\exp(i\pi\theta Y_{\mathsf{s}_{k+1, \uparrow}}X_{\mathsf{s}_{k, \uparrow}})$ for fixed-point values $\theta<1$ in the $|\theta\rangle$ subspace of the temporary register on top. This is done under catalytic use of a phase-gradient state in register $\mathsf{f}$. The circuit on the right-hand-side features $\pm \pi/4$  rotations of Pauli strings $X_{\mathsf{s}_{k+1,\uparrow}}X_{\mathsf{s}_{k,\uparrow}}$, controlled flips of all qubits in $\mathsf{f}$ denoted by the gates `X', and a controlled Gidney adder~\cite{Gidney2017Adders}, which adds the binary number stored in the register under the `$+$' gate into the register under the `=' gate. For the Givens rotation uncomputations, this adder is replaced with an in-place subtractor, symbolized by `$-$'.}
    \label{fig:givens}
\end{figure}
\noindent

\subsection{Givens rotations}
\label{givensrotsection}
The $\mathsf{Select}$ circuit is based on a subroutine that we shall call $\mathsf{subselect}_i$, performing the operation,
\begin{align}
\label{eq:basisrotatedparity}
    \mathsf{subselect}_i\,  |\mu\rangle_{\mathsf{b}_i} \otimes |\psi\rangle_{\mathsf{s}_{\uparrow}} = |\mu\rangle_{\mathsf{b}_i} \otimes U^\dagger_{\mu} Z_{0, \uparrow} U^{\phantom{\dagger}}_{\mu} |\psi\rangle_{\mathsf{s}_{\uparrow}},
\end{align}
defined with respect to one of the $\mathsf{b}_i$-registers. Rather than multiplexing over operators $\lbrace U^\dagger_{\mu} Z_{0, \uparrow} U^{\phantom{\dagger}}_{\mu}\rbrace$ directly, $\mathsf{subselect}_i$ multiplexes over a set of parameters that define $U_\mu$: it loads a series of angles $\lbrace\theta_{\mu k}\rbrace_k$  into temporary registers based on the index $\mu$. The quantum circuit for the $\mathsf{subselect}_i$ routine is depicted in Figure~\ref{fig:givens}$(a)$. Let us say that each angle $\theta_{\mu k}$ is represented in $\beth+1$ bits, i{.}e{.} we would store a positive number $\theta<1$ in its binary fixed-point form to represent an angle $2\pi \theta$. These angles then become inputs for programmable Givens rotations, i.e.\ circuits that implement Givens rotations
$G_k(\theta)$ for numbers $|\theta\rangle$ stored in an input register: $|\theta\rangle |\psi\rangle \mapsto |\theta\rangle G_k(\theta) |\psi\rangle $. Programmable Givens rotations make catalytic use of phase gradient states, 
\begin{align}
    |\Psi\rangle = \frac{1}{\sqrt{2^{\beth+1}}}\sum_{k=0}^{2^{\beth + 1}-1} e^{i\pi k / 2^{\beth}} |k\rangle\, ,
\end{align}
that we store in a register $\mathsf{f}$. With such a state $|\Psi\rangle$, as well as a temporary register holding an angle $\theta$,  we can perform the phase operation $\mathcal{A}|\theta\rangle|\Psi\rangle= \exp(-i2\pi \theta)|\theta\rangle|\Psi\rangle$ with an in-place adder $\mathcal{A}$, adding the content of the temporary register into $|\Psi\rangle$. Using this phase gradient state addition for rotations $\exp(i2\pi \theta P)$ of some Pauli string $P$, one must additionally flip the sign of the angle, i.e.\ $\theta \mapsto 1 -\theta$, in the $(-1)$ subspace of $P$. This can be achieved by flipping all qubits of $\mathsf{f}$ conditional on the value of $P$~\cite{Gidney2017Adders}. However, a single Givens rotation consists of two (commuting) Pauli string rotations acting on consecutive qubits ($k$ and $k-1$ for $G_k$) with opposite angles:
\begin{align}
G_k(\theta) = \exp\!\left(-i\pi \theta X_{k+1} Y_k \right)\;  \exp\!\left(i\pi \theta Y_{k+1} X_k\right)\, ,
\end{align}
and so $G_k(\theta)$ has the eigenvalues $\lbrace 1, \exp(i2\pi \theta),  \exp(-i2\pi \theta)  \rbrace$. Hence we can implement $G_k(\theta)$  with two separate adders, or we can fuse the two adders into a single circuit: a controlled adder featuring conditional flips of the phase gradient register. The fused-adder circuit, shown in Figure \ref{fig:givens}$(b)$, has a lower AV than the separate-adder circuit, used in~\cite{Lee2020EvenMore}. It even has a lower Toffoli count, considering that the bit strings going into the fused adder correspond to angles $\theta$ rather than $\theta/2$, and are therefore one bit shorter.
Putting everything together, the $\mathsf{subselect}_i$ operator multiplexes over all possible values $|\mu\rangle$ of the register $\mathsf{b}_i$, writing the respective angles $\theta_{\mu k}$ into temporary registers and feeding those into a sequence of programmable Givens rotations circuits. 
While the set of angles  $\lbrace \theta_{\mu k}\rbrace_k$  constructs the operators $U_\mu$ associated with the 2-body terms, we also define the set of angles $\lbrace \phi_{\mu k}\rbrace_k$ used to construct the 1-body operators $V_\mu$:
\begin{align}
\label{eq:linops}
    \prod_{k=0}^{N/2-2} G_{k}\!\left(\theta_{\mu k}\right) = U_\mu\, , \qquad \prod_{k=0}^{N/2-2} G_{k}\!\left(\phi_{\mu k}\right) = V_\mu\, .
\end{align}
The relationship between the rotation angles $\theta_{\mu k}$ and the tensors $\chi^\mu_p$ are discussed in Appendix~\ref{rounding}.
For the 2-body terms, $\mathsf{Select}$ must call two instances of $\mathsf{subselect}_i$ sequentially, one for $\mathsf{b}_i=\mathsf{b}_0$ and the other on $\mathsf{b}_i=\mathsf{b}_1$. 
\subsection{Select}
\label{selectsection}
Besides multiplexing over different Givens rotations, the  $\mathsf{Select}$ circuit in  Figure \ref{fig:select}$(a)$ fulfills a number of tasks.  It needs to:
\begin{itemize}
    \item[(i)] Switch between one- and two-body terms;
    \item[(ii)] Swap registers $\mathsf{s}_\uparrow$ and $\mathsf{s}_\downarrow$ to account for spins $\sigma$, $\tau$ as in Eq.~\eqref{eq:biham};
    \item[(iii)] Only keep the terms $\sigma\neq\tau$ in the case $\mu=\nu$;
    \item[(iv)] Access the cases $\mu>\nu$ not encoded in $|\Lambda\rangle$; and
    \item[(v)] Make sure the whole $\mathsf{Select}$ circuit is self-inverse.
\end{itemize}
To accomplish task $\mathrm{(i)}$, we use the qubit $\mathsf{c}$ as an indicator for when to load the angles $\phi_{\mu k}$ rather than $\theta_{\mu k}$. The tasks (ii) and (iii) are fulfilled using the qubits $\mathsf{x}$, $\mathsf{y}$ and $\mathsf{z}$. 
Identifying the spins $\uparrow$, $\downarrow$ with bit values $0$, $1$, the qubits $\mathsf{y}$ and $\mathsf{z}$ are prepared in $\propto\sum_{\sigma}\sum_{\tau(\sigma)} |\sigma\rangle|\tau\rangle$, and qubit $\mathsf{x}$ acts as a switch for this state to be either $|+\rangle|+\rangle$ or $(|01\rangle + |10\rangle)/\sqrt{2}$. For task $\mathrm{(iv)}$,   $\mathsf{b}_0$ and $\mathsf{b}_1$ are swapped controlled on the subspace of qubit $\mathsf{i}$. The $\mathsf{Select}$ is self-inverse following the argument outlined in~\citet{Lee2020EvenMore}, where it is suggested that most of what we see in Figure $\ref{fig:select}$ is just a unitary transform of the actual $\mathsf{Select}$ circuit, the  (self-inverse) Pauli $X$ operator applied to qubit $\mathsf{i}$ towards the end.\\

\begin{figure*}[tb]
    \centering
    \includegraphics[width=\linewidth]{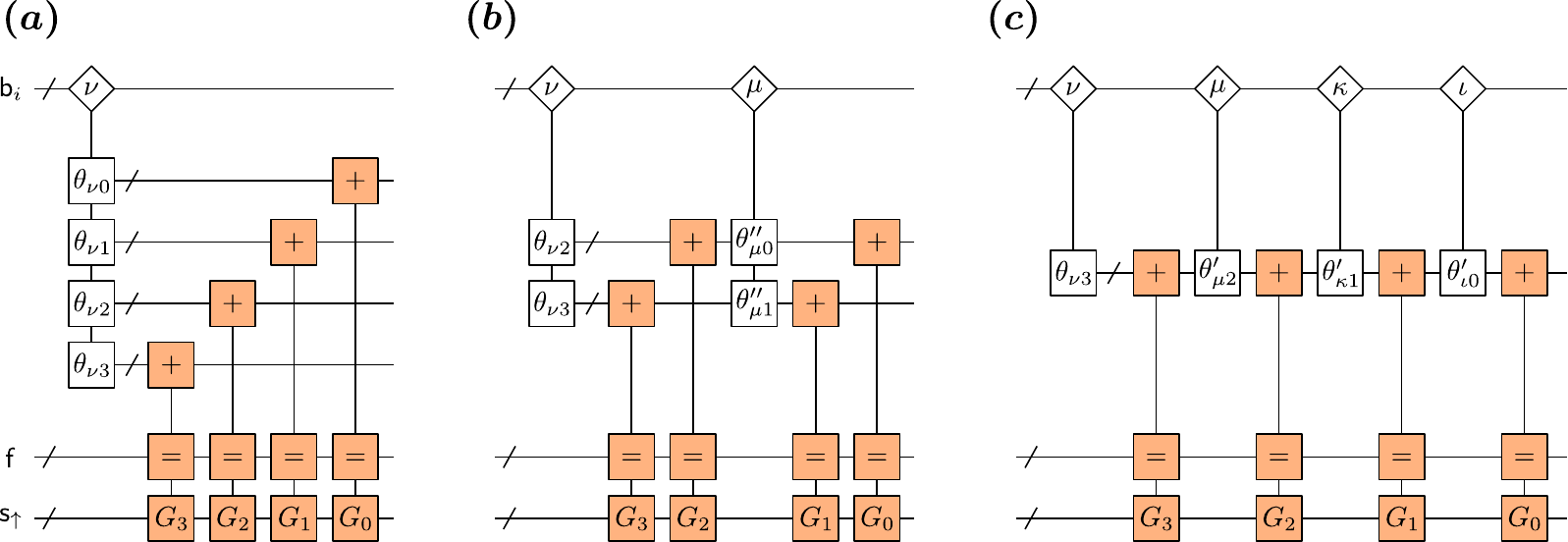}
    \caption{Batching the angle loading exemplified on a part of the $\mathsf{Select}$ circuit. $\boldsymbol{(a)}$ No batching: all angles are loaded at the same time, each into a separate temporary quantum register. Givens rotations then feed on those individual registers. This results in the version of $\mathsf{Select}$ depicted in Figure~\ref{fig:select}. It has the lowest magic state requirements but uses the largest number of qubits. $\boldsymbol{(b)}$ Some batching: a number of angles loaded at the same time, right before their use in Givens rotation circuits. This reduces the qubit cost but requires repeated data loading, increasing the Active Volume. After the Givens rotation circuits of one batch are done, new angles are loaded into the already-allocated temporary registers. Those registers still hold the values $|\theta_{\nu 2}\rangle$ and $|\theta_{\nu 3}\rangle$, but the data strings $\theta^{\prime\prime}_{\mu 0}$ and $\theta^{\prime\prime}_{\mu 1}$ are defined such overwriting the temporary registers leaves them in the states $|\theta_{\nu 0}\rangle$ and $|\theta_{\nu 1}\rangle$.  $\boldsymbol{(c)}$ The maximum amount of batching: each angle is loaded right before its respective Givens rotation circuit. This version uses only one temporary register to store angles. After the $k$-th Givens rotation circuit, we are turning the state $|\theta_{\nu k}\rangle$  into $|\theta_{\nu (k+1)}\rangle$ by loading the data strings $\theta^\prime_{\nu {(k+1)}}$.}
    \label{fig:batching}
\end{figure*}

We have two more comments to make about $\mathsf{Select}$. First, while the angle parameters $\theta_{\mu k}$ and $\phi_{\mu k}$ are fixed-point numbers represented with $\beth + 1$ bits of precision, we have found that they can be represented with only $\beth$ bits, as we have the freedom to set their first bit to zero.  To our knowledge, this has not been considered in prior art. Our definition of $\beth$ is hence somewhat different from the work of Lee \emph{et al.}  For more details, we would like to refer the reader to  Appendix~\ref{rounding}. Second, note that the $\mathsf{Select}$ circuit in Figure~\ref{fig:select} is optimized for the lowest magic state count and uses a lot of auxiliary qubits. Specifically, the elbow dataloader needs to reserve $\beth$ bits of precision for each of the  $N/2 - 1$ angles, leading to an overhead of 741 qubits for P450 ($N=116$ and $\beth=13$). This qubit requirement can be alleviated by a simple modification: instead of loading all $N/2-1$ angles at once, one may decide to load only $r$ angles at a time, using $r \times \beth $ auxiliary qubits. After one batch of angles has been fed into $r$ Givens rotations, a subsequent dataloader would load the next batch into the same temporary registers, overwriting their previous values. This process is depicted in Figure~\ref{fig:batching}. Batching in groups of $r$ angles would require $(N-2)/r$ more dataloaders per $\mathsf{subselect}$, which increases its overall complexity regardless of how we measure it, but it can prove to be useful when the number of qubits is a limiting factor.

\section{Framework for hardware and runtime calculations}
\label{sec:hardware}
\subsection{Active Volume and physical resource estimation}
\label{avsection}

This subsection outlines our methodology for estimating the physical resources required to execute a quantum computation using photonic fusion-based quantum computing hardware. We consider two architectures from the literature: a baseline (BL) interleaved fusion-based quantum computing (FBQC) architecture~\cite{bombin2021interleaving} and the Active Volume (AV) architecture~\cite{litinski2022active}. To simplify our resource estimation, we will assume our architectures implement surface codes and logical operations are carried out using lattice surgery~\cite{horsman2012surface}.

We will first quantify the logical resources, and then translate those estimates into physical resource counts. For both of the aforementioned architectures, the canonical metric for quantifying logical resources is the \emph{spacetime volume}. Roughly speaking, one can calculate the spacetime volume by multiplying the number of logical qubits by the number of logical cycles required to complete the computation. A \emph{logical cycle} comprises $d$ \emph{code cycles}, where $d$ refers to the code distance and a code cycle is the time required to measure every syndrome. We often measure time in units of logical cycles because a logical cycle is the period in which a single lattice surgery operation is implemented~\cite{litinski2019game}.

As the spacetime volume is a product of the two main resources used in a computation -- number of qubits and time -- tradeoffs between these two resources are captured well by this metric. The quantum architecture determines the layout of the logical qubits and how logical operations are implemented, so the choice of architecture can have a profound influence on the spacetime volume. Thus, our first task will be to quantify the spacetime volume for both the baseline and AV architectures.

\begin{figure*}
    \centering
    \includegraphics[width=0.6\linewidth]{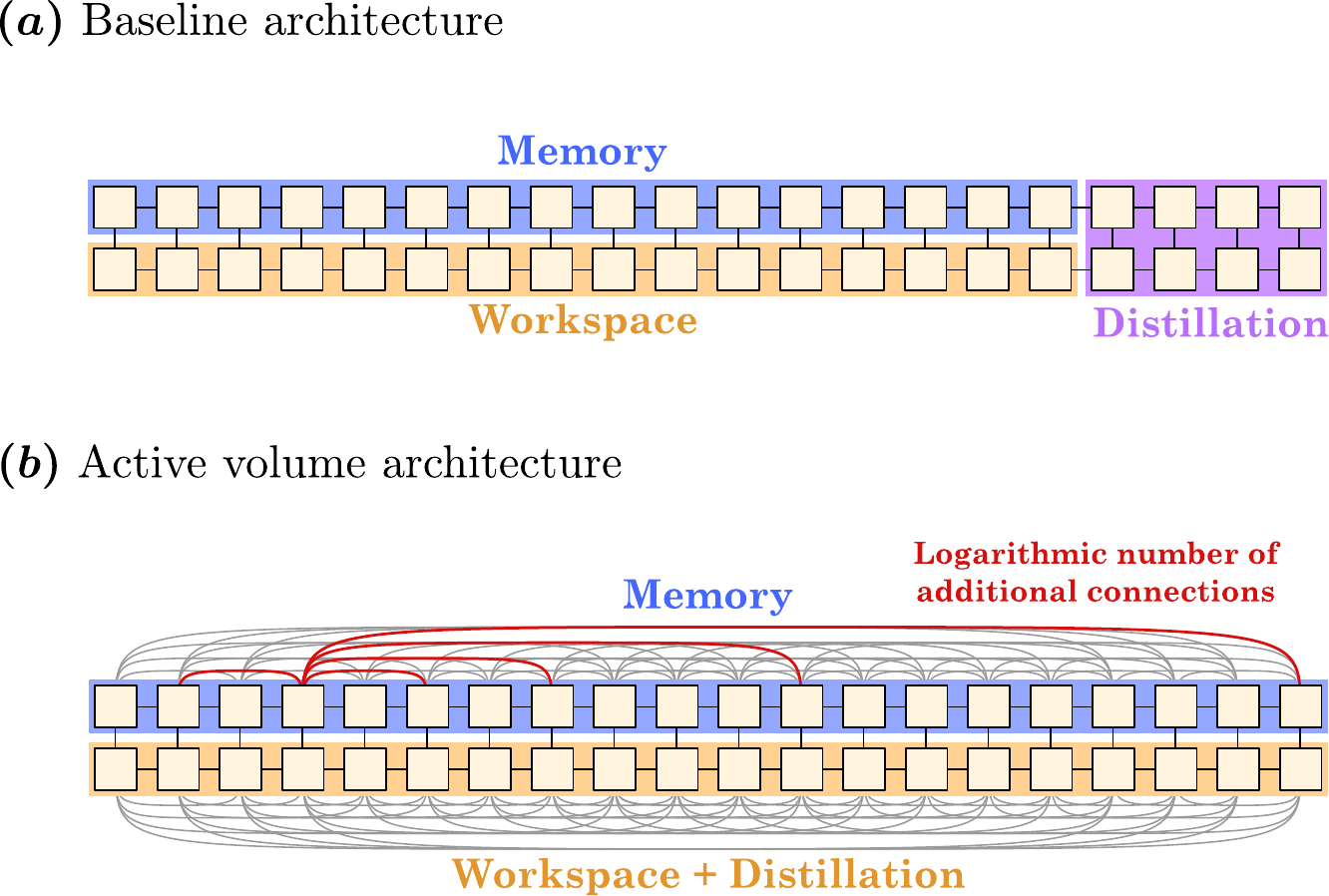}
    \caption{FBQC architectures considered in this work, where nodes correspond to surface code patches. $\boldsymbol{(a)}$ Baseline architecture, with connectors signifying geometric adjacency, enabling lattice surgery operations, between workspace (yellow), memory (blue), and distillation (purple) qubits. $\boldsymbol{(b)}$ Active Volume architecture, in which distillation is done with the workspace qubits. Within the memory and workspace groups, a logarithmic number of additional connections per qubit lets us swap qubits quickly. These additional connections allow workspace qubits to connect with any memory qubit with a small number of quickswap operations. Thus, unlike the baseline architecture, we can vary the number of workspace qubits while maintaining connectivity.}
    \label{fig:archs}
\end{figure*}

The baseline architecture assumes $2m$ logical qubits, of which half are memory qubits (we will in fact use $m$ as the number of memory qubits) and the other half are workspace qubits. In addition to memory and workspace qubits, a third group of qubits is reserved for distilling magic states to implement T gates, as shown in Fig.~\ref{fig:archs}$(a)$. In the simplest variant of this architecture, we use enough magic state factories to produce 1 T gate per logical cycle.~\footnote{We also assume that each individual workspace qubit is used infrequently enough that we can rotate rough and smooth edges to face the workspace qubits when each one is needed.} Then while the next T gate is being produced, the workspace qubits can consume the magic state produced in the previous logical cycle. This simple production and consumption strategy ensures that the total computation time only depends on the number of T gates that have to be produced. In fact, we can use the total T count, or $n_T$, as a representative proxy for the time dimension of the spacetime volume metric. Thus, the spacetime volume of the BL architecture is $2 m \times n_T$, where $m \times n_T$ is also referred to as the \emph{circuit volume}.

In the AV architecture, circuit volume is replaced by Active Volume. Active Volume is simply the number of lattice surgery operations used in the computation. The Active Volume architecture has a total of $n = m + w$ logical qubits: $m$ qubits are allocated for memory, and $w$ qubits are allocated for workspace. Unlike in the BL architecture described above, the number of workspace qubits is not restricted to be equal to the number of memory qubits and we have the option to vary the number of workspace qubits~\footnote{In the architecture presented in the original AV paper the authors assumed that $m = w$. This was largely for simplicity.}. The workspace qubits carry out magic state distillations and logical operations so the more workspace qubits we have, the faster we will compute. This architecture has a logarithmic number of non-local connections among memory qubits and workspace qubits that facilitate operations such as quickswaps to help ``pack'' each logical cycle with lattice surgery operations, see Figure~\ref{fig:archs}$(b)$. To emphasize that we are ``packing'' lattice surgery operations as much as possible, we often refer to lattice surgery operations in the AV architecture as~\emph{logical blocks}. `Logical blocks' is thus the unit of AV and circuit volume counts.
Optimistically, the additional connections given in the AV architecture allow us to execute $w$ logical blocks in each logical cycle. A quantum computation with a total Active Volume of $b_{\text{av}}$ logical blocks results in an estimated spacetime volume of $(w + m) \times (b_{\text{av}} / w) = (1 + \frac{m}{w})b_{\text{av}}$. To roughly estimate $b_{\text{av}}$ or the total Active Volume of an algorithm, one could look up and sum the numbers of logical blocks in an Active Volume architecture for various operations and subroutines from Table 1 of Ref.~\cite{litinski2022active}.

In summary, for a BL architecture, the spacetime volume is approximately 2 $\times$ circuit volume, while for an AV architecture, the spacetime volume is approximately $(n/w)$ $\times$ Active Volume:
\begin{widetext}
\begin{align}
    \text{Spacetime volume (STVol)} \; &= \; \text{number of logical qubits} \times \textrm{number of logical cycles} \notag \\
    \text{STVol}_{\text{BL}} &= \underbrace{2m}_\textrm{logical qubits} \times \underbrace{n_T}_\textrm{logical cycles} = \ 2 \times \underbrace{m \times n_T}_{\mathclap{\textrm{circuit volume}}} \label{eq:stvol_BL}\\ 
    \text{STVol}_{\text{AV}} &= \underbrace{(m + w)}_\textrm{logical qubits} \times \underbrace{b_{\text{av}}/w}_\textrm{logical cycles} = \ \frac{n}{w} \times \underbrace{b_{\text{av}}\vphantom{\sum}}_{\mathclap{\text{Active Volume}}} \; = \;  \frac{n}{w} \ \Bigg( \sum_{i \in \textrm{subroutines}} b_{\text{av},i} \Bigg)\label{eq:stvol_AV},
\end{align}
\end{widetext}
where $b_{\mathrm{av},i}$ is the number of logical blocks in each subroutine used in the computation. Due to the additional operations that pack or parallelize more logical blocks per logical cycle, an Active Volume architecture can significantly reduce the overall spacetime volume of a given computation, implying that Active Volume can often be significantly lower than circuit volume~\cite{litinski2022active}. 

Now that we have estimated the logical resources, we can calculate the physical resources required for executing the computation, including runtime and \emph{footprint} \footnote{We occasionally refer to the total number of interleaving modules as footprint, as it determines the physical size of the quantum computer and due to the term `size' being inaccurate. If the number of IMs is to increase, do we need to build a larger machine or do we need to make the quantum computer smaller? We say that devices with more (less) interleaving modules have a bigger (smaller) footprint.}, using additional information about the hardware.
For a general surface code-based quantum computation, the physical runtime can be estimated as,
\begin{align}
    t_{\textrm{comp}} = \textrm{number of code cycles} \times \textrm{code cycle time},
\end{align}
where the number of code cycles is the number of logical cycles times the code distance $d$. For the baseline architecture, we recall that the number of logical cycles is $n_T$. For the Active Volume architecture, the number of logical cycles is $b_{\text{av}}/w$, where $b_{\text{av}}$ is the total Active Volume. The code cycle time depends on the hardware. For instance, the Google Quantum AI team often reports the code cycle time of their superconducting circuit-based quantum computers to be on the order of microseconds~\cite{fowler2012surface, gidney2021factor, Goings2022P450, acharya2024quantum}. The physical footprint can be estimated based on the number of logical qubits and the code distance.

We now specify the equations for estimating the runtime and footprint for photonic FBQC hardware.
At a high level, FBQC involves the generation of entangled few-photon resource states, storing them in optical fiber or delay lines, and implementing entangling two-photon measurements between different resource states that are called fusions~\cite{bombin2021interleaving}. The generation and measurement of resource states are prescribed by a \emph{fusion graph}, which can be mapped from a spacetime diagram, a common representation used to specify a surface-code-based quantum computation. This mapping determines fusions by corresponding stabilizer measurements in the spacetime diagram representation. 

A FBQC has a number of \emph{Interleaving Modules} (IMs), which generate and hold these photonic resource states. Active Volume architectures use a modified version of the interleaving modules in baseline architectures, due to the different connectivity requirements (see Fig.~4 in~\cite{bombin2021interleaving} and Fig.~23 in~\cite{litinski2022active}), but we shall regard the two types as comparable. These resource states pass through delay lines connecting interleaving modules over specific durations before participating in measurements called fusions. In this simplified picture, FBQC can be viewed as a game of delaying and routing resource states before fusions -- the length of the delay line determines the ratio of logical qubits per IM. In contrast, connections between different interleaving modules and fusions facilitate operations.

The expression to estimate the footprint can be more easily understood by considering how logical qubits are realized in FBQC. The number of logical qubits depends on the maximum number of resource states present simultaneously in the interleaving module, which is linearly dependent on the length of the longest delay line in the interleaving module $l_\textrm{delay}$ and the number of interleaving modules $n_{\mathrm{IM}}$. Assuming a surface code, each logical qubit requires a total of $d^2$ resource states, despite practical implementations of surface code typically requiring $2d^2-1$ physical qubits. This is because the fusions between resource states act as syndrome measurements, eliminating half the qubits that would typically be needed to hold the syndrome data in a surface code~\cite{bartolucci2023fusion}.

Thus, the number of logical qubits $n$ in a FBQC at any given time can be written as
\begin{align}
\label{eq:av_num_logical_qubits}
    n = \underbrace{n_{\textrm{IM}} \, n_{\textrm{RS per IM}}}_{\substack{\textrm{number of resource states}\\ \text{over all IMs}}} \times \underbrace{1/d^{2}}_{\substack{\text{ratio of logical qubits}\\ \text{to resource states}}}
\end{align}
where $n_{\textrm{RS per IM}}$  is the number of resource states that can be stored in one interleaving module. We can write $n_{\textrm{RS per IM}}$ as
\begin{align}
    n_{\textrm{RS per IM}} = r_\textrm{IM} \times \underbrace{\frac{l_\textrm{delay}}{c_\textrm{fiber}}}_{\mathclap{\text{delay time}}}
\end{align}
where $r_\textrm{IM} \approx 10^{9}$ Hz~\cite{litinski2022active} is the rate at which an IM produces resource states, $l_{\textrm{delay}}$ is the length of the fiber optic cable, and $c_\textrm{fiber} \approx 2 \times 10^5$ km/sec is the speed of light in fiber optic cables.  
The footprint can then be estimated by rearranging the equation with respect to the number of IMs:
\begin{align}\label{eq:IM_count}
    n_{\textrm{IM}} = \frac{n d^2 c_\textrm{fiber}}{r_\textrm{IM}l_\textrm{delay}}.
\end{align}
While our explanation derives the number of IMs based on the memory requirement of a computation, in practice, the number of IMs is fixed.

If we denote the total number of resource states needed to complete the computation as $\textrm{Vol}_{\textrm{comp}}$, then the runtime can be estimated by the time required to produce all of those states:
\begin{align}
\begin{split}\label{eq:physical_runtime}
    t_{\textrm{comp}} &= \textrm{Vol}_{\textrm{comp}} \times \underbrace{ \left(n_{\textrm{IM}} r_{\textrm{IM}}\right)^{-1}}_{\mathclap{\substack{\text{resource states created}\\ \text{in 1 second}}}}\\ 
    &= \frac{\textrm{Vol}_{\textrm{comp}} \, l_{\textrm{delay}} }{n d^2\,c_\textrm{fiber}}.
\end{split}
\end{align}
This volume can be estimated based on the spacetime volume, which is defined by the architecture
\begin{align}
\label{eq:comp_volume}
    \textrm{Vol}_{\textrm{comp}} = \textrm{STVol} \times d^3.
\end{align}
where $\textrm{STVol}$ is the spacetime volume as given by Eqs.~\eqref{eq:stvol_BL} and~\eqref{eq:stvol_AV}. Putting the last three equations together with Eqs.~\eqref{eq:stvol_BL} and~\eqref{eq:stvol_AV}, we can calculate the total time for both the baseline and Active Volume architectures as
\begin{align}
    t_{\textrm{comp}} = \frac{l_{\textrm{delay}} \, d}{c_\textrm{fiber}}\quad \times \quad \left\lbrace\begin{tabular}{ll} $n_T \quad$ &for BL architectures  \\ \\  $b_{\text{av}}/w$ &for AV architectures.
    \end{tabular} \right. 
    \label{eq:time-av}
\end{align}
 Although these equations may seem remarkably similar, we will see in the next section that optimizing these two equations to reduce the runtime can be quite different.

\subsection{Architecture Optimization}
\label{subsec:archopt}
Eqs.~\eqref{eq:IM_count} and~\eqref{eq:physical_runtime} show a linear spacetime tradeoff between the number of IMs and the physical runtime in the baseline architecture. If all the interleaving modules use a shorter delay line length, 
Eq.~\eqref{eq:physical_runtime} dictates that this will proportionally reduce the runtime of the algorithm. However, in order to run the algorithm we must still have $n$ qubits available to satisfy memory requirements. By Eq.~\eqref{eq:IM_count} we must increase the number of interleaving modules $n_\textrm{IM}$ to make up for the decrease in qubits incurred by decreasing $l_\textrm{delay}$. One can see this as an example of a spacetime tradeoff where we can decrease the time of the computation by increasing the size of the computer. This spacetime tradeoff is a feature of FBQC and is referred to as \emph{Interleaving} \cite{bombin2021interleaving}.

In the Active Volume architecture, we have access to another tradeoff worth considering: trading code distance for the number of workspace qubits. 
If we can decrease the distance and increase the number of logical qubits, those new logical qubits can be added to the workspace and reduce the number of logical cycles. 

To demonstrate this tradeoff mathematically, let us explicitly calculate the runtime for the case of the AV architecture by combining Eqs.~\eqref{eq:IM_count} and~\eqref{eq:time-av}. We start with a quick rearrangement of Eq.~\eqref{eq:time-av} and then plug in Eq.~\eqref{eq:IM_count} for $n$
\begin{align}
    t_{\textrm{comp}}^\textrm{AV} 
    &= \frac{l_{\textrm{delay}} }{c_\textrm{fiber}} \frac{b_{\text{av}}\, d}{n - m}\\
    &= \frac{b_{\text{av}}\, d^3}{n_{\textrm{IM}} r_\textrm{IM} - \frac{c_\textrm{fiber}}{l_\textrm{delay}}md^2}.\label{eq:explicit_comp_time}
\end{align}
Note that decreasing $d$ will decrease the computation time and, in contrast to the baseline architecture, increasing $l_\textrm{delay}$ also decreases the computation time. Lowering the distance decreases the time because it lowers the time it takes to complete a logical cycle. Increasing $l_\textrm{delay}$ decreases the computation time because the qubits made available by increasing the delay length can be used to execute logical blocks. However, one cannot simply reduce the distance of the code to achieve an arbitrarily low runtime. The distance has to be kept large enough such that the computation does not incur an error. We derive this minimal distance $d_\textrm{min}$.

It is common to model the probability of an error in a computation using an exponential model~\cite{Goings2022P450, acharya2024quantum}:
\begin{align}
    \textrm{probability of failure per logical block} = 10^{-\alpha \frac{d}{2}}.
\end{align}
The parameter $\alpha$ is defined by
\begin{align}
\label{eq:alpha}
    \alpha = \log_{10}\left(\frac{p_\textrm{phys}}{p_\textrm{thresh}}\right)
\end{align}
where $p_\textrm{phys}$ is the physical error rate and $p_\textrm{thresh}$ is the threshold error rate of the quantum error correction code. 
To account for varying levels of hardware performance affecting $p_\textrm{phys}$, in the analysis in Section~\ref{sec:results} we report numbers for both $\alpha=1$ and $\alpha=1/2$. We expect that $\alpha = 1/2$ is a more realistic estimate for an early FTQC device. 
However, previous literature \cite{Goings2022P450} has used $\alpha = 1$ for their resource estimates. However, since current developments~\cite{acharya2024quantum} are at the level of $\alpha=0.33$, we shall regard the $\alpha=1$ case as optimistic and treat the case $\alpha=0.5$ as realistic alternative.

We can use this model to relate the distance $d$ to the number of logical blocks that can be executed:
\begin{align}
\label{eq:computable_blocks1}
    p_{\mathrm{f}} &= 1 - (1 - 10^{-\alpha \frac{d}{2}})^{\text{STVol}}\\
    &\approx \text{STVol} \times  10^{-\alpha \frac{d}{2}}\\
    \text{STVol} &\approx p_{\mathrm{f}} 10^{\alpha \frac{d}{2}},
\end{align}
where $p_{\mathrm{f}}$ is the probability of failure for the overall algorithm and $\text{STVol}$ is the total spacetime volume of the computation in logical blocks. The above equation is valid for both Active Volume and circuit volume: $\text{STVol}$ can be identified with $\text{STVol}_\text{AV}$ or $\text{STVol}_{\text{BL}}$. We can directly relate this to the Active Volume by using Eq.~\eqref{eq:stvol_AV}
\begin{align}
\label{eq:computable_blocks2}
    b_{\text{av}} &\approx \frac{w}{n} p_{\mathrm{f}} 10^{\alpha \frac{d}{2}}.
\end{align}
Finally, let us use the total number of qubits $n=w+m$ and Eq.~\eqref{eq:IM_count} to give
\begin{align}
\label{eq:computable_blocks3}
    b_{\text{av}} &\approx \left(1 - m\frac{d^2 c_\textrm{fiber}}{n_{\textrm{IM}} r_\textrm{IM}l_\textrm{delay}}\right)  p_{\mathrm{f}} 10^{\alpha \frac{d}{2}}.
\end{align}
The above equation tells us the maximum number of logical blocks $b_{\text{av}}$ we can execute with $m$ memory qubits and a total probability of failure $p_{\mathrm{f}}$. We can then define $d_\textrm{min}$ to be the minimal distance, which satisfies the above relation. There is no closed-form solution for $d_\textrm{min}$, however, since $d_\textrm{min}$ will almost certainly be less than 100, it can be found numerically. By inserting this $d_\textrm{min}$ into Eq.~\eqref{eq:explicit_comp_time} we obtain an explicit relation for the minimal computational time.

\section{Results}
\label{sec:results}
In this section, we present logical and physical resource estimates, as well as runtimes of the electronic structure calculation for the P450 cytochrome molecular benchmark system. As in all former studies we take the heme group from the active site of P450 as a model system for the enzymatically relevant part of the protein and reduce the number of orbitals in the Hamiltonian by selecting a (63e, 58o) active space model in the high-spin ($S=5/2$) electron configuration~\cite{Goings2022P450,p450zenodo}. 
In Section~\ref{hamsection}, we discuss the BLISS-THC factorization of P450~\cite{blissthczenodo}, obtaining not just the new 1-norm $\tilde{\lambda}_\mathrm{THC}$ that gives us an immediate speedup over THC, but also a suitable BLISS-THC rank $M$ as well as the bit precision parameters $\aleph$ and $\beth$. Those numbers are the basis for the circuit compilation Section~\ref{avresultsection}, where we present logical resource estimates along with the combined speedup for BLISS-THC, AV compilation and circuit modifications. In Section~\ref{runtimeresultsection}, we estimate wallclock runtimes using various amounts of interleaving. Due to optimizations of the code distance, the speedup is increased even more. We conclude Section~\ref{runtimeresultsection} with a discussion about the minimal number of IMs for a target runtime.  A final attempt to reach a larger speedup is made in Section~\ref{batchingsection}, where we discuss how loading angles in batches within the block encoding influences the runtime of the entire calculation. 

The analysis that we have subjected P450 to can of course also be applied to other molecular systems. The interested reader may find results on the speedup for the electronic structure calculations of FeMoco in Appendix~\ref{sec:femoco}.

\subsection{BLISS-THC factorization}
\label{hamsection}
Just like other factorization schemes, BLISS-THC requires a user-specified error threshold to determine the optimal rank and bit precision requirements for the coefficients $\tilde{\zeta}_{\mu\nu}$ as well as the rotation angles associated with the unitary implementation of $\chi^\mu_p$ \cite{Lee2020EvenMore,Goings2022P450}. However, practical reasons demand that the optimal two-index tensors $\tilde{\zeta}_{\mu\nu}$ and $\chi^\mu_p$ must be found to arbitrary precision first. To that end, we perform a numerical optimization, for which  we propose the BLISS-THC cost function,
\begin{widetext}
\begin{align}
    C(\alpha,\beta,\chi,\tilde{\zeta}) = \frac{1}{2}\sum_{pqrs}\left[g^{\mathrm{(BI)}}_{pqrs}(\alpha,\beta) - \sum_{\mu\nu} \tilde{\zeta}_{\mu\nu} \chi^\mu_p \chi^\mu_q  \chi_r^\nu \chi_s^\nu\right]^2 + \rho \left( \sum_k |\tilde{t}_k| + \tfrac{1}{2}\sum_{\mu\nu}|\tilde{\zeta}_{\mu\nu}| - \tfrac{1}{4}\sum_{\mu}|\tilde{\zeta}_{\mu\mu}| \right),
    \label{BLISS_THC_Cost}
\end{align}
\end{widetext}
as a modification to the regularized cost function of Goings \emph{et al.}~\cite{Goings2022P450}, which includes prior knowledge of the computational cost of quantum phase estimation via the 1-norm weighted by the regularizer, $\rho$. While the original cost function was only optimized for the tensors $\zeta_{\mu\nu}$ and $\chi^\mu_p$, we include here the symmetry-shift and BLISS coefficients, denoted by $\alpha$ and $\beta$ respectively. The numerical optimization of the THC tensors attempts to find an optimal tradeoff between minimizing the $\ell_2$-norm with the smallest THC rank $M$ while also minimizing the 1-norm. Unlike previous work, we explicitly include the 1-body contribution to the 1-norm in the regularization term. Fundamentally, the need for this inclusion arises from the BLISS coefficients, $\beta_{pq}$, which affect 1-body and 2-body Hamiltonian operators. Since this is BLISS-THC, the conventional four-index tensor $g_{pqrs}$ is replaced with the block-invariant tensor $g^{\mathrm{(BI)}}_{pqrs}(\alpha,\beta)$ which is implicitly defined with respect to coefficients $\alpha$ and $\beta$. Similar to prior THC workflows \cite{Goings2022P450}, our optimization procedure starts with a symmetric canonical polyadic decomposition of the Cholesky vectors to provide the initial guess, which takes around 6 minutes running on a CPU. This is followed by a regularized optimization of the BLISS-THC tensors, carried out with 15 thousand L-BFGS-B iterations on a single Nvidia GeForce RTX 4090 consumer-level GPU, which takes another six minutes

The cost function of Eq.~\eqref{BLISS_THC_Cost} evaluates the $\ell_2$-norm error of the truncated BLISS-THC Hamiltonian in Eq.~\eqref{BI_Hamiltonian} with respect to the exact Hamiltonian of Eq.~\eqref{electronic_hamiltonian}. While this is a suitable error metric for the optimization procedure, it does not provide a reliable proxy to the correlation energy error. We
therefore estimate the truncation error $\epsilon_{\text{trunc.}}$ via CCSD(T) calculations based on the reconstructed 1-electron and 2-electron integrals using PySCF~\cite{sun2020pyscf}. From the tabulated results, we choose the final BLISS-THC rank $M$, satisfying the user-specified threshold $\epsilon_{\text{trunc.}} \leq 0.3$ mEh similar to~\cite{Lee2020EvenMore}. An expanded discussion of these salient points is provided in Appendix \ref{rounding}. 

We present the analysis of the THC rank with respect to CCSD(T) values for $\epsilon_{\mathrm{trunc.}}$ in Table~\ref{tab:sample-data}. Notably, we observe an increase of the 1-norm as a function of the THC rank while still maintaining a sub-milliHartree CCSD(T) correlation energy error for nearly all cases. These results establish that the lowest possible rank is $M = 160$, which meets the target error threshold of $0.3\,\mathrm{mE_h}$. 

\begin{table}[tb]
\centering
\caption{BLISS-THC results for the factorization rank $M$, the $\ell_2$ norm of the difference tensor $g_{pqrs}-\widetilde{g}_{pqrs}$ (where $\widetilde{g}_{pqrs}$ is the version of $g_{pqrs}$, reconstructed from the factorization), the CCSD(T) error and the BLISS-THC 1-norm $\tilde{\lambda}_{\mathrm{THC}}$. We ultimately choose the THC rank $M=160$ (highlighted in blue) due its CCSD(T) being below the threshold of 0.3 $\mathrm{mE_h}$. Recall that Goings \emph{et al.} find a rank of $M=320$, which is exactly twice the rank for BLISS-THC, as well as a 1-norm of $388.9\,\mathrm{E_h}$, roughly 3 times as high as ours.} \vspace{0.5cm}
\label{tab:sample-data}
\begin{tabular}{
    c  
    S[table-format=1.3]  
    S[table-format=-1.3]  
    S[table-format=3]    
}
\toprule
\multicolumn{1}{c}{\makecell{Rank  \\ $M$}} &
\multicolumn{1}{c}{\makecell{$\ell_2$-error \\   ($\mathrm{E_h}$)}} &
\multicolumn{1}{c}{\makecell{CCSD(T) \\ error  ($\mathrm{mE_h}$)}} &
\multicolumn{1}{c}{\makecell{1-norm \\ ($\mathrm{E_h}$)}} \\
\midrule
120 & 0.309 &-2.14 & 132.3  \\ 
\addlinespace[1pt]
140 & 0.219 & -1.15 & 132.5  \\  
\addlinespace[1pt]
\color{blue}160 & \color{blue}0.202 & \color{blue}0.25 & \color{blue}130.9   \\ 
\addlinespace[1pt]
180 & 0.122 & 0.11 & 134.4  \\ 
\addlinespace[1pt]
200 & 0.106 & -0.03 & 133.3  \\ 
\addlinespace[1pt]
220 & 0.065 & 0.32 & 139.3  \\ 
\addlinespace[1pt]
240 & 0.054 & 0.25 & 137.4  \\
\addlinespace[1pt]
260 & 0.042 & 0.29 & 138.1  \\
\addlinespace[1pt]
280 & 0.033 & 0.26 & 138.9  \\
\addlinespace[1pt]
300 & 0.032 & -0.04 & 138.9  \\
\addlinespace[1pt]
320 & 0.024 & 0.28 & 139.1 \\
\bottomrule
\end{tabular}
\end{table}

\begin{figure}[tb]
    \centering
    \includegraphics[width=\linewidth]{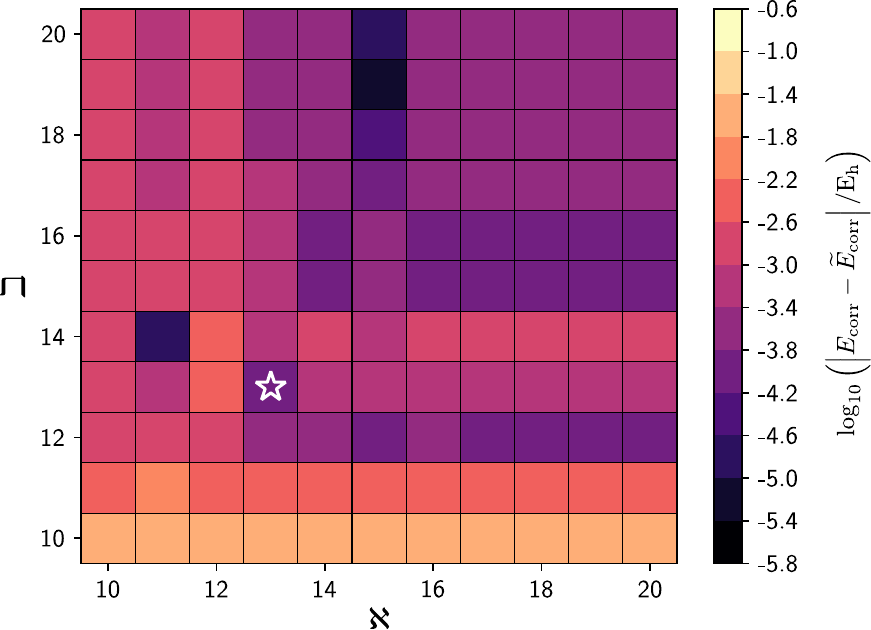}
    \caption{Heat map of the error between the CCSD(T) correlation energy calculated with the uncompressed Hamiltonian and the BLISS-THC compressed Hamiltonian as a function of fixed point precisions $\aleph$ and $\beth$. The star indicates the combination $(\aleph,\beth)=(13,13)$ selected for the BLISS-THC of P450. We have also performed a DMRG calculation with bond dimension $1500$ for the tuples $(\aleph,\beth)=(13,13)$ and $(\aleph,\beth)=(11,14)$, confirming an error in the correlation energy of $0.08\,\mathrm{mE_h}$ and $0.10\,\mathrm{mE_h}$ respectively. For comparison, Goings \emph{et al.}\ \cite{Goings2022P450} have found a tuple of $(\aleph, \beth) = (10, 18)$ with respect to our definition of $\beth$, however, we believe that their report has neglected the preparation of 1-body terms, which would have changed the landscape of both $\aleph$ and $\beth$.}
    \label{fig:heatmap}
\end{figure}

With the rank now fixed, we set out to find the number of bits for the precision of the Alias Sampling state preparation and Givens rotations, $\aleph$ and $\beth$. In Figure~\ref{fig:heatmap}, we present a 2D heat map of CCSD(T) correlation energy error defined with respect to  $\aleph$ and $\beth$. Similar to prior art, we observe a non-trivial oscillatory behavior in the CCSD(T) correlation energy error. To adhere to the error threshold of 0.3 mEh, we select the bit precision parameters $(\aleph,\beth)=(13,13)$. This choice is confirmed by the DMRG results, where the error to the correlation energy introduced by the BLISS-THC procedure is $0.08\,\mathrm{mE_h}$ for a bond dimension of 1500. For details on the DMRG results, we would like to refer the reader to Appendix~\ref{sec:dmrg}.

\subsection{From factorization to Active Volume, circuit volume and speedups}
\label{avresultsection}
\begin{table*}[tb]
\centering
\caption{Toffoli count, memory qubit highwater, circuit volume as well as Active Volume for the electronic structure calculation of P450 with respect to the THC Hamiltonian of~\cite{Lee2020EvenMore} and the BLISS-THC Hamiltonian developed in this work. For both Hamiltonians, we offer the choice between the $\mathsf{Select}$ circuit of~\cite{Lee2020EvenMore} and our modified $\mathsf{Select}$ in Figure~\ref{fig:select}. In this table, we have highlighted the circuit volume $b_{\mathrm{cv},0}$ with respect to the prior-art $\mathsf{Select}$~\cite{Lee2020EvenMore} and the THC Hamiltonian of Goings \emph{et al.}\ \cite{Goings2022P450}, in blue, and $b_{\text{av},1}$, the value for the Active Volume of our modified $\mathsf{Select}$ using the BLISS-THC Hamiltonian, has been highlighted in orange. The ratio  $b_{\mathrm{cv},0} / b_{\text{av},1}$  is important for the computation of the relative speedup factor in Eq.~\eqref{eq:volratio}. Note that we have effectively re-estimated the logical resources for the circuit of Goings \emph{et al.}\ with respect to their reported rank, 1-norm, and precision parameters.}\vspace{.5cm}
\begin{tabular}{l@{\hskip 10pt}c@{\hskip 10pt}r@{\hskip 10pt}r@{\hskip 10pt}r@{\hskip 10pt}r}
\toprule
Hamiltonian & Circuit mods & Toffolis  & Memory & Circuit Volume & Active Volume  \\
 \midrule
\multirow[c]{2}{*}{THC} & no & $7.786 \times 10^9$ & 1,357 & \cellcolor{blue!15} $42.264 \times 10^{12}$ & $1.678 \times 10^{12}$ \\
 & yes & $7.647 \times 10^9$ & 1,298 & $39.703 \times 10^{12}$ & $1.594  \times 10^{12}$ \\
 \midrule
\multirow[c]{2}{*}{BLISS-THC} & no & $1.761 \times 10^9$ & 1,058 & $7.450 \times 10^{12}$ & $0.249 \times 10^{12}$ \\
  &  yes & $1.714 \times 10^9$ & 999 & $6.848 \times 10^{12}$ & \cellcolor{orange!20} $0.228 \times 10^{12}$ \\
\bottomrule
\end{tabular}
\label{tab:volumes}
\end{table*}
With all parameters of the factorization fixed, we can now compile the quantum phase estimation routine central to this electronic structure calculation, while neglecting any overhead that might be incurred in the initial state preparation, as well as repetitions of the entire algorithm to increase the statistical confidence of the results. The computational volume of the routine, as well as its qubit \emph{highwater} \footnote{We will occasionally refer to  the number of memory qubits as highwater, due to its non-additivity: the highwater is determined by the largest number of qubits used in all the subroutines of the algorithm. The term is mostly used when resources are estimated, {i.}{e.} when we are trying to determine the size of the memory register.}, will critically depend on which $\mathsf{Select}$ circuit we pick, and we have several versions to choose from: the $\mathsf{Select}$  developed in Section~\ref{blockencodingsection} differs somewhat from prior art~\cite{Lee2020EvenMore} due to our use of fused adders,  tighter analysis for the bit precision, and the treatment of diagonal terms in the THC matrix. Also, the possibility of batching offers a number of variations on each version of $\mathsf{Select}$. For now, let us focus on versions of the circuit where all angles are loaded in a single batch: that is, the $\mathsf{Select}$ with our modifications as well as the literature version for comparison. To correct some inaccuracies in the logical resource counts of the literature reference, we have re-estimated the resources of the THC algorithm with respect to the parameters $\aleph=10$, $\beth=18$ \footnote{This is indeed the value of prior art after our redefinition of $\beth$.}, $\lambda = 388.9\,\mathrm{E_h}$ and $M=320$ from \cite{Goings2022P450}.

From only their logical resource estimates, we can derive relative speedups between two quantum calculations without explicitly computing their respective runtimes. For the total runtime speedup, taking into account the effects of circuit modifications, BLISS-THC and AV compilation, the following two quantum programs are relevant:  computation 1 is our version of the algorithm run on an AV architecture using BLISS-THC, and computation 0 is the literature version of the electronic structure calculation for reference. Let us say we run both computations on  devices with the same footprint, which for now shall mean identical code distances, interleaving lengths and qubit counts. The latter would pose a condition to be enforced. Following Eq.~\eqref{eq:time-av}, the speedup $t_{\textrm{comp}, 0} / t_{\textrm{comp}, 1}$ with respect to the individual runtimes $t_{\textrm{comp}, i}$ of computations $i=0,1$  is
\begin{align}
\label{eq:speedup}
    \frac{t_{\textrm{comp}, 0}}{t_{\textrm{comp}, 1}} =  \frac{n_{T,0}\, w_1}{b_{\text{av},1}}\, ,
\end{align}
 where  $b_{\text{av},1}$ and $w_1$ are the AV count and workspace size of computation 1, and $n_{T,0}$ is the T count of computation 0. 
Due to being run in a baseline architecture, computation 0 has a 1:1 ratio between the sizes of workspace and memory, while the number of workspace qubits can be freely chosen for computation 1. Through circuit modifications and BLISS, the memory of computation 1 will be somewhat relaxed, and so we reassign the freed-up memory qubits to the workspace, such that the sum of memory and workspace qubits remains constant for both computations. To obtain a lower bound on the runtime speedup, we will not re-assign the distillation qubits from computation 0 in computation 1.  With $w_1=2m_0-m_1$, where $m_0$ and $m_1$ are the memory qubit highwaters of computation 0 and 1, respectively, the speedup in Eq.~\eqref{eq:speedup}  becomes
\begin{align}
\label{eq:volratio}
        \frac{t_{\textrm{comp}, 0}}{t_{\textrm{comp}, 1}}  \; = \; \left(2 - \frac{m_1}{m_0}{}\right) \frac{b_{\mathrm{cv},0}}{b_{\text{av},1}}\, ,
\end{align}
where $b_{\mathrm{cv},0}$ is the circuit volume of computation 0.
While circuit volume is easily accounted for, we must obtain an upper bound on the total Active Volume by adding the AV costs of the algorithm's sub-components. A repository for the Active Volume costs of all subroutines relevant to this case can be found in Table 1 of~\cite{litinski2022active}. We present the results of our logical resource estimations for computations featuring THC and BLISS-THC Hamiltonians and with and without our modifications to $\mathsf{Select}$ in Table~\ref{tab:volumes}. Plugging the highlighted values for circuit volume and  Active Volume into Eq.~\eqref{eq:volratio}, it becomes apparent that we have achieved a total speedup of $\Mark{233.9\times}$ over the  runtime  of the THC algorithm on a baseline architecture.  While our modifications to the $\mathsf{Select}$ circuit have a small, yet positive effect on the AV, it is not the only way in which they reduce the runtime. Through the modifications, $\mathsf{Select}$ also calls fewer qubits, making room for a larger workspace. This reduction in the memory is due to encoding each of the 57 Givens rotation angles using $\beth$ rather than $\beth + 1$ qubits. An even larger reduction in  memory is achieved by switching from THC to the BLISS-THC Hamiltonian, where $\beth$ is \Mark{13} rather than 18.
 While $\beth$ has the biggest impact on the global qubit highwater, the $\mathsf{Prepare}$ circuit allocates roughly $M \sqrt{\aleph/2 + \log M}$ qubits for a QROAM dataloader in Alias Sampling. This local qubit highwater is almost on par with the one of $\mathsf{Select}$. 

\begin{figure*}[tb]
    \centering
    \includegraphics[width=\linewidth]{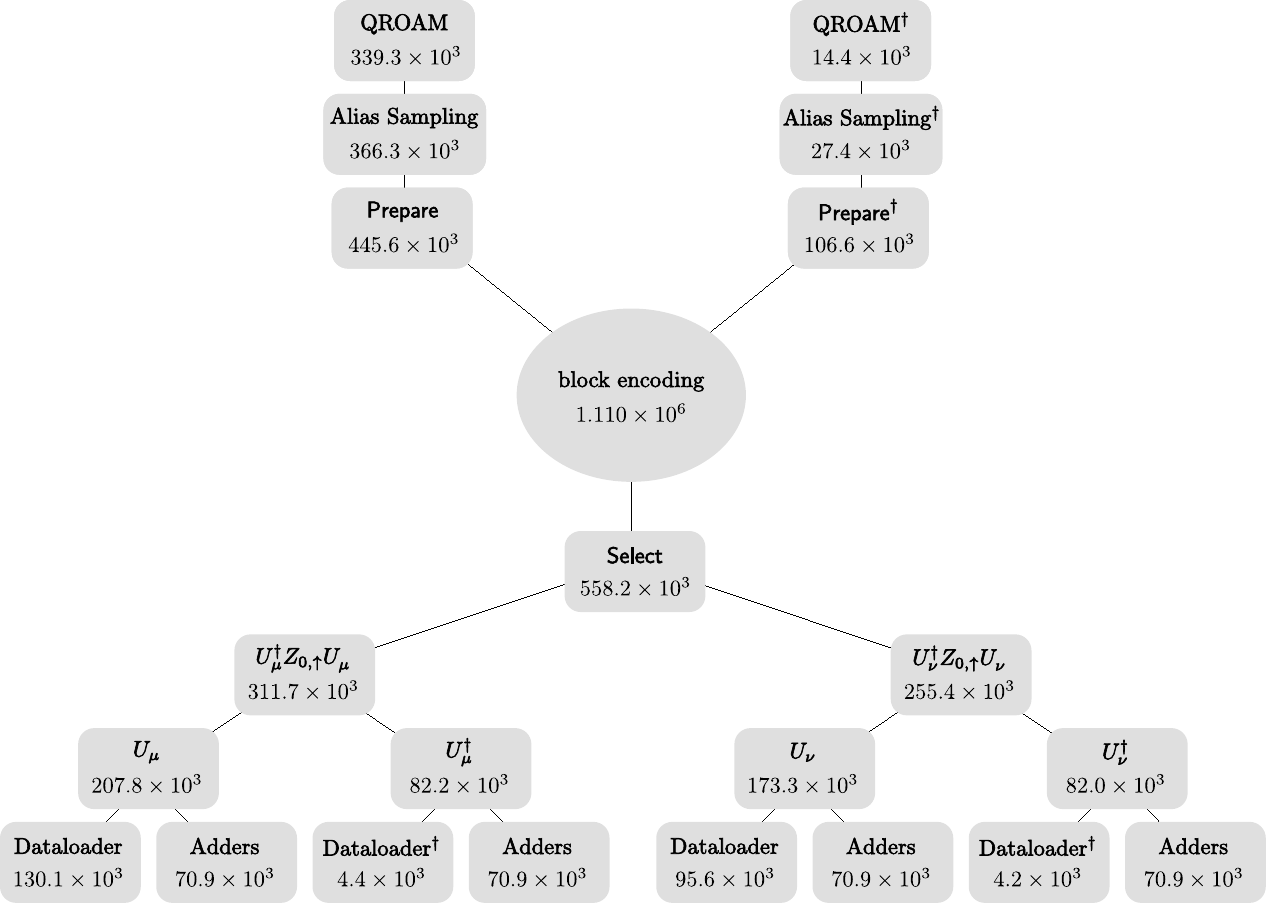}
    \label{fig:callgraph_av}
    \caption{A callgraph showing the AV for a hierarchy of relevant subroutines emanating from the BLISS-THC block encoding of P450 (center). The subroutines labeled $\boldsymbol{U_\mu^\dagger Z_{0,\uparrow} \!\!U_\mu^{\phantom{\dagger}}}$ and $\boldsymbol{U_\nu^\dagger Z_{0,\uparrow}\!\! U_\nu^{\phantom{\dagger}}}$ are versions of $\mathsf{subselect}$, multiplexing over indices $\mu$ and $\nu$, where the former also the multiplexes over 1-body operators. The basis rotations $U_\mu$ and $U_\nu$ include Givens rotations and dataloaders.}
\end{figure*}

Before we move on to runtimes, we present a breakdown of AV costs between the algorithmic subroutines. The callgraph in Fig.~\ref{fig:callgraph_av} provides insight into the relative cost between the different parts of a single BLISS-THC block encoding. In computation 1  $\Mark{40}\%$ of the total AV is spend on $\mathsf{Prepare}$ and $\Mark{10}\%$ on  $\mathsf{Prepare}^\dagger$, while the $\mathsf{Select}$ routine accounts for the remaining $\Mark{50}\%$. Inverse routines are usually cheaper, as they tend to feature measurement-based uncomputation, such as in right elbows and inverse dataloaders. It is worth noting that AV compilation shifts the relative costs of some subroutines: in previous cost metrics, the adders associated with the Givens rotations account for roughly $4 N \beth$ Toffoli gates in a single instance of $\mathsf{Select}$, which dwarves the roughly $2M+N$ Toffoli gates in the dataloaders by an order of magnitude. However, the AV costs of adders and dataloaders are quite similar. This is likely due to the cost associated with the Clifford gates: writing and overwriting $(2M + N)(N/2-1)$ different angles of $\beth$ bits into temporary quantum registers requires many conditional bit-flips.

\subsection{From computational volume to runtimes and device footprint}

\begin{table*}[tb]
\centering
\caption{Numbers of interleaving modules, code distances and runtimes for the electronic structure calculation of P450 as a function of the delay length ($l_\textrm{delay}$) as a tunable parameter for interleaving. The table considers two computations 1) the BLISS-THC calculation with circuit modifications run on an AV architecture; and 2) the  THC calculation without circuit modifications run on a BL architecture. For a direct comparison, the number of IMs for the AV calculation is matched to the IM requirements of the BL run, and code distances are computed for a logical error probability of $\leq10\%$  with respect to the whole algorithm. Additionally, we distinguish the cases of $\alpha=0.5$ and $\alpha=1$. Runtimes are presented in the format hours~:~minutes~:~seconds.}\vspace{.5cm}

\begin{tabular}{l@{\hskip 15pt}c@{\hskip 15pt}c@{\hskip 15pt}rr@{\hskip 15pt}rr@{\hskip 15pt}rr@{\hskip 15pt}rr}
\toprule
 & & & \multicolumn{2}{r}{1m delay}\hspace{15pt} & \multicolumn{2}{r}{10m delay}\hspace{15pt} & \multicolumn{2}{r}{100m delay}\hspace{15pt} & \multicolumn{2}{r}{1000m delay}\hspace{15pt} \\
 Algorithm (Architecture)& $\alpha$ & $d$ & IMs  & time & IMs & time& IMs  & time & IMs  & time \\
\midrule
THC (BL) & 0.5 & 60 & 1,954,080  & 2:35:44 & 195,408  & 25:57:16 & 19,541  & 259:32:38 & 1,955 & 2595:26:18 \\
BLISS-THC (AV) & 0.5 & 50& 1,954,080  & 0:00:20 & 195,408  & 0:03:17 & 19,541  & 0:32:42 & 1,955 & 5:26:47 \\
\midrule
THC (BL) & 1 & 30& 488,520  & 1:17:52 & 48,852  & 12:58:38 & 4,886  & 129:46:19 & 489  & 1297:43:09 \\
BLISS-THC (AV) & 1 & 25 & 488,520  & 0:00:10 & 48,852 & 0:01:39 & 4,886  & 0:16:21 & 489  & 2:43:17 \\
\bottomrule
\end{tabular}

\label{tab:runtimes}
\end{table*}
\label{runtimeresultsection}
For Eq.~\eqref{eq:volratio}, we have assumed the code distances of both computations to be the same, but we can do better: due to its lower computational volume, the AV-based BLISS-THC computation requires a lower code distance. While this would give the runtime an additional boost according to  Eq.~\eqref{eq:time-av}, lowering the distance of every logical qubit would also reduce the required footprint. Following Section~\ref{subsec:archopt}, we minimize the code distance and compute the runtimes after assigning freed-up resources to the workspace. In doing so, we can ensure a direct comparison between the two different computations, by keeping the number of IMs fixed. Note that the runtime improvement of $\Mark{233\times}$,  obtained in the previous subsection, is retained as lower bound on the speedup, independent on the physical error rate and the threshold. Both of these parameters are captured in the error suppression constant $\alpha$ of Eq.~\eqref{eq:alpha}. In Goings \emph{et al.}'s paper, runtimes are obtained for $\alpha=1$; an optimistic assumption for early fault-tolerant quantum computers, that we would like to contrast with a more realistic scenario of $\alpha=0.5$.

In Table~\ref{tab:runtimes}, we finally present runtimes, IM numbers and code distances for the two relevant computations: the THC algorithm of~\cite{Lee2020EvenMore} run on a baseline architecture, as well as our BLISS-THC algorithm run on an AV architecture. 

For both computations, we keep the number of IMs fixed, and distinguish the optimistic scenario of $\alpha=1$ from the realistic scenario of $\alpha=0.5$. In all cases, the total wallclock speedup is roughly $\Mark{476\times}$. Using the length of the optical fiber to tune the amount of interleaving, footprints are traded off against runtimes for the computations considered. The shorter the delay length, the lower the ratio of resource states per interleaving module, which makes the computation faster, but increases the number of interleaving modules required.

When we want to keep the quantum computer as small as possible in physical size, the delay lengths need to be maximized. To still reduce the runtime, we can then change the ratio of workspace to memory qubits: the more qubits we dedicate to the workspace, the more logical blocks can be executed in parallel. 

In the following example, we want to compute the minimum number of IMs required to run our BLISS-THC algorithm in a specific time frame. Let us say we fix the interleaving length to 2km -- a large value at which we start to notice the effects of fiber loss~\cite{bombin2021interleaving}. As the number of memory qubits is fixed by the algorithm, the minimum number of IMs is determined by how many qubits we will need in the workspace. In Table~\ref{tab:imnumbers}, we provide the results for a runtime of 73 hours, which is the runtime reported by Goings \emph{et al.}~\cite{Goings2022P450} for an algorithm with $\Mark{4.6 \times 10^6}$ physical qubits.  We also consider the runtime of 1 hour as an intermediate milestone on the way to industrial utility. For both time scales, Table~\ref{tab:imnumbers} presents the minimum number of IMs required under error thresholds parameters of $\alpha = 1$ and $\alpha = 0.5$, along with updated runtimes for the modified $\mathsf{Select}$ operators discussed in the following subsection. 

\subsection{Batching}
\label{batchingsection}
In a final attempt to speed up the computation, we utilize a  tradeoff between qubit highwater and AV counts within the algorithm. Previously, we had decided to only consider versions of $\mathsf{Select}$ that load all Givens rotation angles at once -- those circuits are optimal with respect to Toffoli and AV counts, but have a large memory qubit highwater. In fact, the local highwater of $\mathsf{Select}$ dominates the qubit requirements of the entire algorithm for P450. Allowing circuits that load the angles in batches as depicted in Figure~\ref{fig:batching}, we reduce the qubit requirements of the entire algorithm at the expense of a higher AV. Once the qubits are freed up in memory, they can then be reassigned to the workspace, allowing for a possible runtime speedup assuming the AV hike is overcome in the process.

To avoid confusion, let us refer to any BLISS-THC calculation with a $\mathsf{Select}$ using batched dataloaders as BLISS-THC-b calculation. There is one more modification that we have to make for BLISS-THC-b: in $\mathsf{Prepare}$, the dataloader associated with the Alias Sampling routine, highlighted blue in Figure~\ref{fig:prepare},  needs to be relaxed. By adjusting the lambda-parameter of its LKS construction~\cite{Low2018TradingT}, we can force the QROAM back into a QROM, and reduce the number of qubits allocated by the routine from roughly $M\sqrt{\aleph/2 + \log M}$ to $\aleph + 2\log M$ while approximately doubling its AV. Note that this qubit tradeoff always increases the AV of $\mathsf{Prepare}$.  However, if we did not make this modification, then reducing the local highwater of $\mathsf{Select}$ would at some point have no effect on the global highwater, as the qubits freed up in $\mathsf{Select}$ would still be needed in $\mathsf{Prepare}$.

\begin{figure}[tb]
    \centering
    \includegraphics[width=\linewidth]{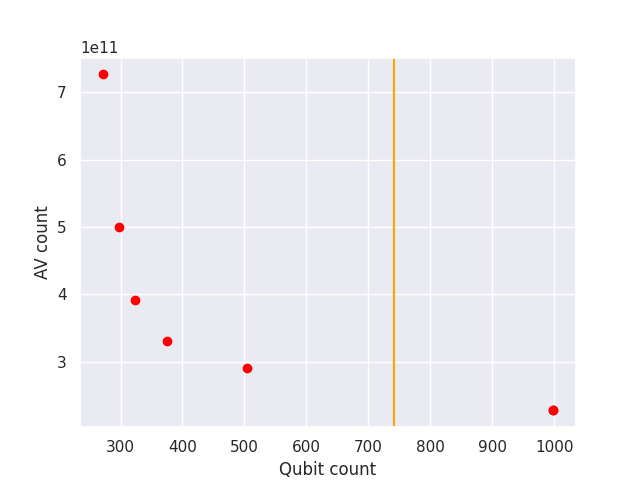}
    \caption{Active Volume and logic qubit count of  BLISS-THC-b for various batch sizes. BLISS-THC-b is a version of our algorithm for BLISS-THC, where the angles for the programmable Givens rotations are loaded in batches as indicated in Figure~\ref{fig:batching}. As an additional modification, we turn the QROAM~\cite{Low2018TradingT}  in $\mathsf{Prepare}$ back into a QROM. If we did not do this, we would not be able to access the qubit numbers left of the orange separator, as the reduction in the qubit count of $\mathsf{Select}$ would be over-towered by the local qubit highwater of $\mathsf{Prepare}$ and all total qubit counts left of the orange line would lie on the line instead. While clearly not optimal with respect to  magic-state counts, this modification has a smaller impact on the AV, which we can now trade off for the number of logical qubits. The lowest number of qubits achievable is $\Mark{271}$, at an AV of approximately $\Mark{ 7.26 \times 10^{11}}$ logical blocks for the algorithm in which we load every angle before the corresponding Givens rotation, as displayed in Figure \ref{fig:batching}$(c)$. The degree of batching is not made explicit here, but higher qubit numbers correspond to larger batches, and the right-most point represent the standard BLISS-THC version, where all angles are loaded at once.}
    \label{fig:avbatching}
\end{figure}

The AV numbers for various instances of BLISS-THC-b with different batch sizes are presented in Figure \ref{fig:avbatching}, showing also the qubit minimum that a magic-state-optimal QROAM in Alias Sampling would have imposed. 

From this plot, we chose the instance of BLISS-THC-b that minimizes the runtime. Table~\ref{tab:imnumbers} contains not just the number of IMs for targeted runtimes but also provides the lowest runtimes for the corresponding BLISS-THC-b calculations. For the devices with lower IM numbers, we find that BLISS-THC-b is able to provide a speedup, while for the devices with a bigger number of IM, the runtime is unchanged. For the latter, the optimization defaults to the case where all angles are loaded in a single batch, turning BLISS-THC-b back to BLISS-THC as no other number of batches provides a runtime speedup. In the two devices with the smallest IM counts, the optimal runtime is achieved when loading the angles in 7 batches.

As an alternative to batching, we have also considered converting Hamiltonian data into a unary representation, and so relax the angle loading. While one would expect this to come with a large qubit highwater, it is only narrowly beaten by batching, as we discuss in Appendix \ref{unarysection}.

\begin{table}[tb]
    \centering
    \caption{Number of interleaving modules required for corresponding runtimes of BLISS-THC and BLISS-THC-b for four hypothetical devices targeting a $73$-hour and a $1$ hour operation of BLISS-THC at optimistic $(\alpha=1)$ and realistic $(\alpha=1/2)$ assumptions regarding the error model. The 73-hour number is set to match the benchmark of Goings \emph{et al.}~\cite{Goings2022P450}, where the optimistic assumption of $\alpha=1$ is used. Here, the delay length is kept as long as possible at 2km, at which the loss threshold is argued to be only somewhat affected \cite{bombin2021interleaving}. Note that for the smallest device, we cannot achieve 73h directly, so we report the largest runtime below the target.}\vspace{.5cm}
\begin{tabular}{c@{\hskip 15pt}c@{\hskip 15pt}c@{\hskip 15pt}c}
\toprule
\multirow{2}{3em}{$\#$ IMs} & \multirow{2}{1em}{$\alpha$} & \multirow{2}{5.5em}{BLISS-THC runtime (h)}  & \multirow{2}{6.5em}{BLISS-THC-b runtime (h)}\\ \\ \midrule
8184 & 0.5 &  1  & 1 \\
1055 & 1 & 1 & 1 \\
395 & 0.5 & 73 & 41 \\
88 & 1 & 54 & 26 \\
\bottomrule
\end{tabular}
    
    \label{tab:imnumbers}
\end{table}

\section{Conclusion}
\label{sec:conclusion}
New algorithms and methods will find application in an industrial setting only if they can deliver accurate results for systems of relevant sizes in timeframes compatible with industrial development cycles. 
This means that to gauge the utility of quantum computing for industrial applications, actual runtimes must be obtained, not just gate complexities or asymptotic scaling analyses. 
The accurate evaluation of runtimes, however, is a complex task for which many aspects must be considered. First of all, runtime estimates are developed at the focal point of not only quantum algorithms and compilation but also quantum error correction and quantum system design. Suddenly, features like the code threshold, code distance, and logical error rate come into play, along with essential questions about qubit connectivity, magic state distillation protocols, and more. In this work, the above considerations are taken into account, providing a holistic picture for us to present the latest improvements to the electronic structure calculations of molecular systems on the example of the heme group occurring in cytochrome P450. One such improvement is the introduction of BLISS-THC, the new state-of-the-art in a chain of cascading advancements in quantum simulation of molecular systems, and applicable to all quantum architectures. To make BLISS-THC numerically attainable, we have drastically driven down the runtime of the classical pre-processing of all flavors of THC. Factorizing a single P450 Hamiltonian with THC currently takes approximately 6 minutes (plus 6 minutes of optimization warm-up) on a single Nvidia GeForce RTX 4090 consumer-level GPU. As a result, we were able to subject the P450 Hamiltonian to a much tighter factorization procedure, with BLISS-THC not only improving the Hamiltonian 1-norm but also the factorization rank of the tensor. Furthermore, we have reduced the classical precompute so much it has become insignificant compared to prior art \cite{Lee2020EvenMore,Rocca2024}. This cancels a big disadvantage of cost function-based methods compared to diagonalization-based methods like DF. We believe that BLISS-THC, on account of it being a strict improvement over THC in terms of compression, will be beneficial to all kinds of molecular systems, and find applications  beyond electronic structure calculations, such as the  simulation of time dynamics, or the estimation of molecular observables~\cite{Steudtner2023SEVE}.

The most significant contribution to our speedups comes from switching to AV compilation. This mode of running quantum programs is amenable to only some quantum hardware types, and the fusion-based photonic architecture is one of them. However, any platform that allows for flying qubits or nonlocal physical gates could be amenable to Active Volume compilation, and the comparison can be made to a reference architecture with immovable qubits or limited connectivity.  Accounting for different execution rates of logical blocks, even cross-platform comparisons are possible.

With some small algorithmic changes, BLISS-THC and AV, we manage to improve the runtime of the electronic structure calculation by a factor $\Mark{233}\times$ with respect to a reference THC algorithm on the same platform. 

This speedup is agnostic to savings in the code distance, for the account of which we require additional information about the error model. Optimizing the code distance for fusion-based photonic hardware the speedup would even jump up to $\Mark{476\times}$.

We have provided runtimes for several sizes of photonic fusion-based quantum computers, where interleaving is used to trade the runtime against the number of IMs. Note however that we cannot increase the number of IMs indefinitely. When considering larger numbers of IMs, the system may encounter additional bottlenecks not examined in this paper, such as the reaction limit~\cite{Litinski2023_Elliptic}. We leave examination of the high-IM configurations to future work. We also show that qubit tradeoffs between using a single or more batches in the loading of Hamiltonian data for the BLISS-THC algorithm, which conventionally slow down the computation in the baseline architecture, can sometimes be utilized to speed it up.

Some improvements in this work not only lower the computational volume, but also reduce the qubit highwater. With the goal of reducing the wallclock runtime, we have repeatedly converted these space savings into savings of time by enlarging the workspace of the AV architecture. Thus, every resource saved other than time has been put back into the device, just so we can have a larger speedup over the algorithms used in prior art. However, perhaps we should also consider device footprints. Many tradeoffs in this work could be used to fit the electronic structure calculation on a quantum computer smaller than the baseline architecture required for the reference computation. 
In that spirit, we have derived minimal device footprints for computations of a fixed duration, by varying the number of workspace qubits. Even larger space savings can be achieved by changing the algorithm: the original THC routine requires a number of auxiliary qubits that exceeds by far the number of qubits necessary to represent the system. While BLISS-THC gives us an immediate reduction of \Mark{299} qubits for P450, we have discovered that more substantial space reductions of up to a highwater of only \Mark{271} qubits increase the AV by only a factor of $\Mark{3.2}\times$. While we have shown how this tradeoff could be utilized for runtime reductions, accepting the AV penalty could be reasonable for a smaller quantum computer.

This paper certainly does not mark the last round of runtime improvements for electronic structure computations on a quantum computer. Many promising avenues remain for future speedups. While we have increased the performance of THC with BLISS and reduced the 1-norm of P450 from $389\,\mathrm{E_h}$ to $130.9\,\mathrm{E_h}$, we have not achieved the theoretical 1-norm limit of $69.3\,\mathrm{E_h}$~\cite{Cortes2024}. Other Hamiltonian factorization techniques~\cite{Loaiza2023BLISSIzmaylov, Patel2024_BLISS_DF} might, however, get us closer. Introducing AV is a paradigm shift in algorithmic costing and has led us to change how we think about certain subroutines. In the future, we expect to be able to make profound modifications to quantum circuits, that reflect an optimal choice with respect to the new cost model. For instance, the fact that Clifford gates are now a dominant cost in dataloaders might open a whole new optimization space to be explored. In that spirit, we should also consider costing subroutines like dataloaders with respect to real data rather than considering their worst-case counts. In fact, this approach could go far beyond dataloaders: the runtimes of all subroutines might be reduced by making quantum programs more concrete. What is more, we are currently considering upper bounds on AV of subroutines. A tighter analysis of AV costs would have an immediate impact on projected runtimes without changes to the algorithms.

We are optimistic that the resource estimates in this work will be improved. To this end, we depend on the contributions of many disciplines on all levels of the quantum computing stack. The impact of novel developments in quantum error correction~\cite{Gidney2024Magic,Bravyi2024High,Sahay2023High,Hann2024Hybrid}, for instance, would need to be considered within the cost model. By continuously improving the cost of this use-case, it is our aim to bring practical quantum computing closer and to contribute to the path to industrial value.

\section{Acknowledgments}
AC, CC, WP, SS and MS would like to thank Sam Pallister and Daniel Litinski for the insightful discussions on the subject matter of this project, as well as Sean Greenaway, Sam Morley-Short and others for their advice and support with respect to the utilized resource estimation tools. GLA, MD, NM, RS, MS and CT thank Clemens Utschig-Utschig for his comments
on the manuscript and the support during the project.

\bibliography{paper_main}
\bibliographystyle{apsrev4-2}

\onecolumngrid

\appendix
\section{Speedups for the electronic structure calculation of FeMoco}
\label{sec:femoco}
\begin{figure}[tb]
    \centering
    \includegraphics[width=.55\linewidth]{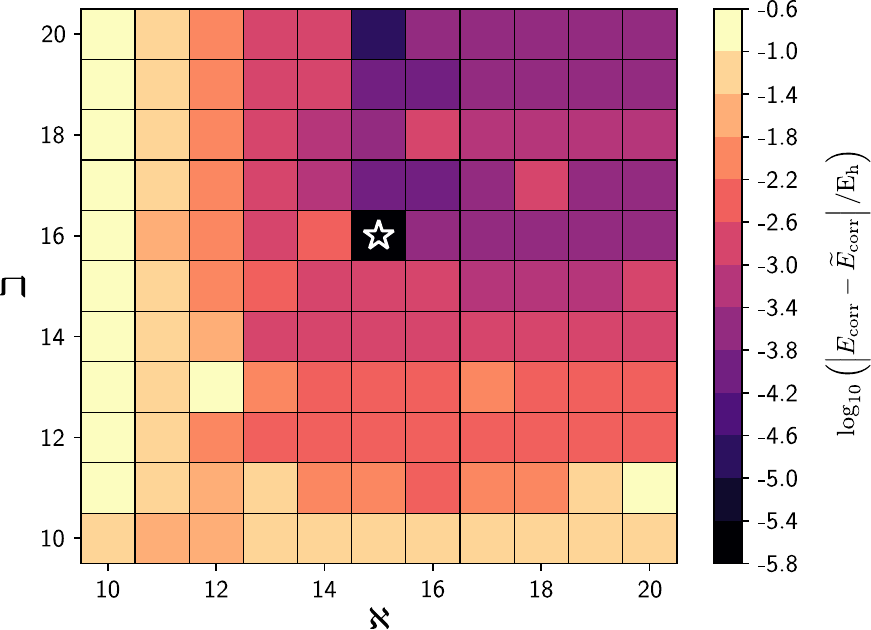}
    \caption{Heat map of the CCSD(T) correlation energy error calculated with the uncompressed Hamiltonian and the BLISS-THC compressed Hamiltonian as a function of fixed point precisions $\aleph$ and $\beth$. The star indicates the final selected combination $(\aleph,\beth)=(15,16)$. For comparison, Lee \emph{et al.} found a tuple of $(\aleph, \beth) = (10, 18)$ with respect to our definition of $\beth$, however, we believe their report neglected the preparation of 1-body terms, which would have changed the CCSD(T) correlation energy error landscape of both $\aleph$ and $\beth$.} 
    \label{fig:heatmap_femoco}
\end{figure}

The iron-molybdenum cofactor, FeMoco (\ce{Fe7MoS9C}), is the active site of nitrogenase, an enzyme playing a crucial role in the biological reduction of nitrogen to ammonia. 
This iron-molybdenum cluster presents a significant challenge for accurate theoretical characterization. Its large size, open-shell electronic configuration requiring a multi-reference description, and the presence of static correlations make conventional computational methods inadequate ~\cite{Montgomery2018FeMocoStrongCorr}. For these reasons, FeMoco is a perfect candidate for quantum chemical calculations on quantum computers, becoming a well-known benchmark system in many studies~\cite{Reiher2017, Lee2020EvenMore, VonBurg2021Catalysis}. In this section, we expand the results presented in the main text with an equivalent analysis for the electronic structure calculation of FeMoco, using the Hamiltonian of Li \emph{et al.}~\cite{li2019electronic}. We obtain equivalent versions of Tables~\ref{tab:speedups}--\ref{tab:volumes} as well as Figure~\ref{fig:heatmap} for FeMoco.
In Table~\ref{tab:sample-data-FeMoco} we report the factorization rank $M$ together with the errors and the 1-norm for the sampling of the eigenspectrum of the FeMoco Hamiltonian. In blue, we highlight the selected factorization which gives the best performance for the target error. Results to the finite-bit analysis of $\aleph$ and $\beth$, are presented as a heatmap in Figure~\ref{fig:heatmap_femoco}. Logical resource estimates, as well as speedups are presented in Tables~\ref{tab:volumes_femoco} and~\ref{tab:speedups_femoco}, where we compare against the re-estimated resources of two factorized Hamiltonians: the original THC version of~\cite{Lee2020EvenMore} and the partially symmetry-shifted THC* version of~\cite{Berry2024Rapid}, finding a speedup of $\Mark{427\times}$ and $\Mark{278\times}$, respectively. A brief summary of all previous and current factorization attempts to the Li Hamiltonian are collated in Table~\ref{tab:factorization_femoco}.

\begin{table}[!b]
\centering
\caption{Logical qubit requirements, Toffoli counts, Active Volume, as well as the 1-norm for electronic structure calculations of FeMoco in a (113e, 76o) active space configuration defined with respect to Sparse, symmetry-shifted Double Factorization (DF*), Tensor Hypercontraction (THC), symmetry-shifted Tensor Hypercontraction (THC*) and BLISS-THC. The Toffoli count of the symmetry-shifted versions of DF and THC (denoted as DF* and THC*) account for the qubitization block encoding cost multiplied by the the total number of repetitions required for the QPE algorithm, taken from Table II in~\cite{Berry2024Rapid}. Note that BLISS-THC approaches 108.9 Eh, the theoretical 1-norm limit for FeMoco~\cite{Cortes2024}, also to within a
factor of 2.} \vspace{0.5cm}
\begin{tabular}{l@{\hskip 10pt}c@{\hskip 10pt}l@{\hskip 10pt}c@{\hskip 10pt}r}
\toprule
\multirow{2}{*}{Factorization} & \multicolumn{4}{c}{FeMoco (113e, 76o) } \\
& Logical qubits & Toffolis & Active Volume & {$\lambda$ ($\mathrm{E_h}$)} \\
\hline
Sparse~\cite{Lee2020EvenMore} & 2,489 & 4,4$\times 10^{10}$ & --- & 1547.3 \\
DF*~\cite{Berry2024Rapid} &  6,402 & 3.2$\times 10^{10}$ & --- & 582.4  \\
THC~\cite{Lee2020EvenMore} & 2,196 & 3.2$\times 10^{10}$   & --- & 1201.5 \\
THC*~\cite{Berry2024Rapid} & 2,194 & 2.1$\times 10^{10}$   &--- & 781.8 \\
THC~[re-estimated for this work] & 2,163 & 3.2$\times 10^{10}$ & $8.8\times 10^{12}$ & 1201.5\\
THC*~[re-estimated for this work] & 2,163 &  2.1$\times 10^{10}$ & $5.7\times 10^{12}$ & 781.8 \\
BLISS-THC [this work] & 1,512 & 4.3$\times 10^{9}$  & $8.4\times 10^{11}$ & 198.9 \\
\bottomrule
\end{tabular}
\label{tab:factorization_femoco}
\end{table}

\begin{table}[th]
\centering
\caption{FeMoco BLISS-THC results as a function of factorization rank $M$, $\ell_2$ error norm, CCSD(T) correlation energy error, and 1-norm. We ultimately choose the THC rank $M=290$ (highlighted in blue) due its CCSD(T) correlation error being below the threshold of 0.3 $\mathrm{mE_h}$. The CCSD(T) calculation uses a high-spin ($S=35/2$) reference using UCCSD(T) in PySCF. For reference, the THC rank $M$ reported in~\cite{Lee2020EvenMore} was equal to $450$. } \vspace{0.5cm}
\label{tab:sample-data-FeMoco}
\begin{tabular}{
    c  
    S[table-format=1.3]  
    S[table-format=-1.3]  
    S[table-format=3]    
}
\toprule
\multicolumn{1}{c}{\makecell{Rank  \\ $M$}} &
\multicolumn{1}{c}{\makecell{$\ell_2$-error \\   ($\mathrm{E_h}$)}} &
\multicolumn{1}{c}{\makecell{CCSD(T) \\ error  ($\mathrm{mE_h}$)}} &
\multicolumn{1}{c}{\makecell{1-norm \\ ($\mathrm{E_h}$)}} \\
\midrule
250 & 0.146 & -0.87& 197.9  \\ 
\addlinespace[1pt]
270 & 0.112 & 0.61 & 198.7   \\ 
\addlinespace[1pt]
\color{blue}290 & \color{blue}0.086 & \color{blue}-0.26 & \color{blue}198.9  \\ 
\addlinespace[1pt]
310 & 0.068 & -0.48 & 198.8  \\ 
\addlinespace[1pt]
330 & 0.056 &-0.26 & 199.6  \\ 
\addlinespace[1pt]
350 & 0.045 & -0.21 & 198.9  \\
\addlinespace[1pt]
370 & 0.039 & 0.09 & 199.0  \\
\addlinespace[1pt]
390 & 0.035 & 0.07 & 202.3  \\
\addlinespace[1pt]
410 & 0.030 & 0.16  & 200.7 \\
\addlinespace[1pt]
430 & 0.025 & -0.20 & 200.8 \\
\addlinespace[1pt]
450 & 0.022 & 0.21 & 202.5 \\
\bottomrule
\end{tabular}
\end{table}

\begin{table}[tb]
\centering
\caption{Toffoli count, number of memory qubits, and circuit/Active Volume for the electronic structure calculation of FeMoco with respect to the $\mathsf{Select}$ circuit of~\cite{Lee2020EvenMore} and our modified $\mathsf{Select}$ in Figure $\ref{fig:select}$, as well as the choice between the THC Hamiltonian of~\cite{Lee2020EvenMore}, the THC* Hamiltonian of \cite{Berry2024Rapid} and novel BLISS-THC Hamiltonian. Note that we have effectively re-estimated the circuit of Lee \emph{et al.}\ \cite{Lee2020EvenMore} with respect to the reported ranks, 1-norms, and precision parameters reported in \cite{Lee2020EvenMore} and \cite{Berry2024Rapid}.}\vspace{.5cm}
\begin{tabular}{l@{\hskip 10pt}c@{\hskip 10pt}r@{\hskip 10pt}r@{\hskip 10pt}r@{\hskip 10pt}r}
\toprule
Hamiltonian & Circuit mods & Toffolis  & Memory & Circuit Volume & Active Volume  \\
 \midrule
\multirow[c]{2}{*}{THC} & no & $31.767 \times 10^9$ & 2163 &  $274.850 \times 10^{12}$ & $8.762\times 10^{12}$ \\
 & yes & $31.201 \times 10^9$ & 2163 & $269.951 \times 10^{12}$ & $8.392  \times 10^{12}$ \\
 \midrule
 
\multirow[c]{2}{*}{THC*} & no & $20.671 \times 10^9$ & 2163 &  $178.841 \times 10^{12}$ & $5.701\times 10^{12}$ \\
 & yes & $20.302 \times 10^9$ & 2163 & $175.654 \times 10^{12}$ & $5.460  \times 10^{12}$ \\
 \midrule
 
\multirow[c]{2}{*}{BLISS-THC} & no & $4.375 \times 10^9$ & 1589 & $27.809 \times 10^{12}$ & $0.888 \times 10^{12}$ \\
  &  yes & $4.282 \times 10^9$ & 1512 & $25.895 \times 10^{12}$ &  $0.837 \times 10^{12}$ \\
\bottomrule
\end{tabular}
\label{tab:volumes_femoco}
\end{table}

\begin{table}[t]
\caption{ Contributions to the speedup of the electronic structure calculation for the Li \emph{et al.} Hamiltonian of FeMoco~\cite{li2019electronic}, as compared to both, the THC-based calculation~\cite{Lee2020EvenMore}, as well as the symmetry-shifted THC*~\cite{Berry2024Rapid}, on a baseline architecture defined in Section~\ref{sec:hardware}.The three steps, 1) Active Volume compilation, 2) the incorporation of BLISS, and 3) improvements to the block encoding circuit, are done in sequence while adding logical qubits saved in the memory to the workspace. The partial speedups reported are improvements to the runtime with respect to the previous step. Multiplying the individual improvements hence results in the total speedup, up to rounding errors: all numbers in this table are rounded down, but the total speedup is computed with the exact estimates. }\vspace{.5cm}
\begin{tabular}{l@{\hskip 10pt}r@{\hskip 10pt}r}
\toprule
& \multicolumn{2}{c}{Speedup {w.}{r.}{t.}} \\
Method & \cite{Lee2020EvenMore} & \cite{Berry2024Rapid} \\
\hline
AV compilation & $31.36\times$ & $31.36\times$ \\ 
THC $\mapsto$ BLISS-THC  &  $12.48\times$ & $8.12\times$\\
Circuit  improvements & $1.09\times$& $1.09\times$\\
\hline
Total: & $427.42\times $ & $278.11\times $\\
\bottomrule
\end{tabular}
\label{tab:speedups_femoco}
\end{table}

\newpage
\section{Error metrics and rounding procedure for approximate tensors}
\label{rounding}
This section describes the algorithm performance model used to choose all algorithm parameters. The overall target accuracy $\epsilon$ is taken to be 0.0016 Hartree (chemical accuracy) for all of the resource estimates in this manuscript. To achieve this target, we consider the total errors given by the sum of the quantum phase estimation error $\epsilon_\mathrm{PEA}$ and Hamiltonian approximation error $\epsilon_\mathrm{THC}$,
\begin{equation}
    \epsilon \geq \epsilon_\mathrm{PEA} + \epsilon_\mathrm{THC}.
\end{equation}
Here, $\epsilon_\mathrm{PEA}$ is the error due to measurement in the phase estimation procedure, including, for instance, spectral leakage effects. In contrast, $\epsilon_\mathrm{THC}$ corresponds to the total approximation and compilation error that arises from directly implementing the THC Hamiltonian on the quantum computer. Note that this error may be bounded as
\begin{align}
    \epsilon_\mathrm{THC} \leq \epsilon_\mathrm{trunc.} + \epsilon_\mathrm{coeff} + \epsilon_\mathrm{rot},
\end{align}
where $\epsilon_\mathrm{trunc.}$ is the truncation error based on the BLISS-THC factorization procedure with a maximum tensor rank, $\epsilon_\mathrm{coeff}$ is the error that arises from coherent Alias Sampling based on implementing the BLISS-THC Hamiltonian coefficients, $\tilde{\zeta}_{\mu\nu}$, with a finite-bit representation, and $\epsilon_\mathrm{rot}$ is the approximation error that arises from implementing the individual Givens rotations needed for $\chi^\mu_p$. 

While strict analytical bounds on the truncation error, coefficient error, and rotation error may be used to estimate $\epsilon_\mathrm{THC}$, these bounds are often quite loose. To circumvent this, we use the procedure that was initially proposed by Lee \emph{et al.}~\cite{Lee2020EvenMore}, where we reconsider the purpose of the quantum phase estimation algorithm as providing an energy estimate of the \emph{correlation} energy, $E_\mathrm{corr}$, defined by the standard ground-state energy $E_\mathrm{G}$ decomposition, $E_\mathrm{G} = E_\mathrm{HF} + E_\mathrm{corr}$, where $E_\mathrm{HF}$ is the Hartree-Fock energy, which can be computed with floating point (64-bit) precision classically. This is quite different from the standard assumption where QPE is considered to provide an estimate of the ground-state energy $E_\mathrm{G}$ with absolute accuracy. Ultimately, this change in mindset relaxes the stringent requirements for the THC rank or bits of precision that would have been necessary otherwise. 

The main challenge with this approach lies in having a reliable assessment of the correlation energy error, which now becomes the proper error metric for $\epsilon_\mathrm{THC}$. In previous work, Lee \emph{et al.}~\cite{Lee2020EvenMore} proposed estimating this error with two separate coupled-cluster calculations with singles, doubles, and perturbative triples (i.e., CCSD(T)). The proposed use of CCSD(T) is due to its broadly recognized status as the ``gold standard'' for computational chemistry. It provides sufficient scalability for large system sizes while preserving size extensivity. In this work, we advocate for CCSD(T) as a scalable approach for providing correlation energy error estimates for the BLISS-THC tensors. Furthermore, we provide numerical validation of the CCSD(T) correlation energy error metric based on DMRG calculations, which we discuss in the results section in greater detail.

\paragraph*{Numerical evaluation of error metrics.} In the following, we provide explicit details on how the BLISS-THC Hamiltonian error metric is computed, considering both rank truncation and bit precision effects. As mentioned in the previous paragraph, we first bound the BLISS-THC Hamiltonian approximation error by the absolute difference between two CCSD(T) calculations,
\begin{equation}
    \epsilon_\mathrm{THC} \geq |E_\mathrm{CCSD(T)}(\mathrm{exact}) - E_\mathrm{CCSD(T)}(M,\aleph,\beth)|,
\end{equation}
where $E_\mathrm{CCSD(T)}(\mathrm{exact})$ corresponds to the correlation energy estimate based on an exact implementation of the 1-body and 2-body integrals with full floating point (64-bit) precision, while $E_\mathrm{CCSD(T)}(M,\aleph,\beth)$ corresponds to the correlation energy estimate based on the reconstructed 1-body and 2-body integrals. The 1-body integrals are reconstructed based on the standard eigen-decomposition of the $T_{pq}$ matrix, while the 2-body integrals are reconstructed based on the BLISS-THC expansion with finite rank, $M$. Finite bit precision requirements are also imposed onto coefficients $t_k$, $\widehat{\zeta}_{\mu\nu}$ and the rotation angles needed to implement the unitary implementation of $\chi^\mu_p$ and eigenvectors of $T_{pq}$. The parameters $\aleph$ and $\beth$ indicate the finite bit precision parameters. Explicitly, the rounding procedure for $\tilde{\zeta}$ is given by,
\begin{align}
    \tilde{\tilde{\zeta}}_{\mu\nu} = 
    \begin{cases}
        u_\zeta^0 \times \text{round}(\tilde{\zeta}_{\mu\nu}/u_\zeta^0 + x)\;\;\text{for}\;\mu\neq\nu \\
        u_\zeta^{\mathrm{d}} \times \text{round}(\tilde{\zeta}_{\mu\nu}/u_\zeta^d + x) \;\;\text{for}\;\mu=\nu\\
    \end{cases}
\end{align}
where $u_\zeta^0 = \tilde{\lambda}_{\mathrm{THC}}/(\mathfrak{d}2^\aleph)$ and $u_\zeta^{\mathrm{d}} = \tilde{\lambda}_{\mathrm{THC}}/(\mathfrak{d}2^{\aleph-2})$ with $\mathfrak{d}=M(M+1)/2+N/2$, and $x$ is a small constant that is used to ensure the proper normalization $\widehat{\zeta}_{\mu\nu}$~\cite{Lee2020EvenMore}. In this work, we parameterize the rotation vector $\vec{\chi^\mu}$ by the set of angles $\lbrace \theta_{\mu p} \rbrace_{p=0}^{N/2-2}$ defined as,
\begin{align}
\label{eq:chi}
    {\chi}^{\mu}_p[\vec{\theta_{\mu}}] =
    \begin{cases}
    \cos( 2\pi \theta_{\mu p})\prod\limits_{q<p}\sin( 2\pi \theta_{\mu q}) \; \; & \text{for}\;p<N/2-1  \\
    \prod\limits_{q=0}^{N/2-2}\sin( 2\pi\theta_{\mu q}) & \text{for}\;p=N/2-1.
    \end{cases}
\end{align}
with
\begin{equation}
    \widetilde{\theta}_{\mu p} = u_\theta \times \text{round}( \theta_{\mu p}/u_\theta ),
\end{equation}
where $u_\theta = 1/2^{\beth + 1}$, meaning the angles are defined with $\beth+1$ bits of precision. However, we can always set the most significant bit to zero, i.e. restrict the angles to $[0, \pi)$. This parametrization corresponds to the standard  Euclidean space mapping of a multidimensional spherical coordinate system. The absolute value of all $\chi^\mu_{p}$ can be set with angles in the range of $[0, \pi/2]$ where all sines and cosines are positive. The larger range of $[0, \pi]$ allows cosines to be negative, and so all $\chi^\mu_{p}$ for $p<N/2-1$ may acquire a sign. The only coordinate that would need the full range of $[0, 2\pi)$ to be negative is $\chi^\mu_{N/2-1}$, and as we can remove a global minus sign we can always choose $\chi^\mu_{N/2-1}$ to be positive. With a similar argument, we can exclude the point $\pi$ in the range $[0, \pi)$. This parametrization is a departure from the definitions in prior art \cite{VonBurg2021Catalysis, Lee2020EvenMore}. Due to that and another factor of two from equation 49 in \cite{Lee2020EvenMore}, our definition of $\beth$ differs from the one in Lee \emph{et al.}~\cite{Lee2020EvenMore} by $ \beth_{\mathrm{Lee}} = \beth_{\mathrm{here}} + 2$. 
Eq.~\eqref{eq:chi} can then be used to build an approximate representation of the 1-body and 2-body integrals, and the correlation energy $E_\mathrm{CCSD(T)}(M,\aleph,\beth)$ computed.  

\section{DMRG results for P450}
\label{sec:dmrg}
To validate the BLISS-THC truncation parameters, density matrix renormalization group (DMRG) calculations were performed using Block2 based on the reconstructed 1-body and 2-body integrals~\cite{zhai2023block2}. Results are shown in Table \ref{DMRG_results}, highlighting the Hartree-Fock ground-state energy contribution, CCSD(T) correlation energy, and DMRG correlation energy in the last column using a bond dimension of 1500. Interestingly, while the Hartree-Fock energy contribution, and hence the absolute energy, is observed to change up to 0.77$\, \mathrm{E_h}$ in one of the two cases, the correlation energy error computed with both CCSD(T) and DMRG methods remains below the 0.3$\,\mathrm{mE_h}$ threshold for both truncation parameter settings. Moreover, although our DMRG-based results show that the CCSD(T) correlation-energy error remains a good proxy for the P450 benchmark system, further studies on a broader set of benchmarks are needed to determine whether this holds in general or whether alternative metrics should be employed. This question remains open and warrants additional investigation.

\begin{table}[ht]
\centering
\caption{Breakdown of $\ell_2$ error, Hartree-Fock energy, CCSD(T) correlation energy, and DMRG correlation energy for reference P450 active space integrals as well as reconstructed integrals based on $(M,\aleph,\beth)=(160,11,14)$ and $(M,\aleph,\beth)=(160,13,13)$ truncation parameter settings.} \vspace{0.1cm}
\begin{tabular}{c c c c c}
\toprule
Hamiltonian &
\multicolumn{1}{c}{\makecell{$\ell_2$-error \\  ($\mathrm{E_h}$)}} &
\multicolumn{1}{c}{\makecell{$E_\mathrm{ROHF}$ \\  ($\mathrm{E_h}$)}} &
\multicolumn{1}{c}{\makecell{$E_\mathrm{CCSD(T)}-E_\mathrm{ROHF}$ \\  ($\mathrm{E_h}$)}} &
\multicolumn{1}{c}{\makecell{$E_\mathrm{DMRG}-E_\mathrm{ROHF}$ \\  ($\mathrm{E_h}$)}} \\
\midrule
Reference& - & -419.40807 & -0.44864 & -0.45003  \\ 
\addlinespace[1pt]
M=160, ($\aleph,\beth$)=(11,14)  & 0.207 & -418.64088 & -0.44866 & -0.44993  \\  
\addlinespace[1pt]
M=160, ($\aleph,\beth$)=(13,13) & 0.202 & -419.03084 & -0.44872 & -0.44995\\
\bottomrule
\end{tabular}
\label{DMRG_results}
\end{table}

\section{Unary encoding}
\label{unarysection}
\begin{figure}[tb]
    \centering
    \includegraphics[width=0.6\linewidth]{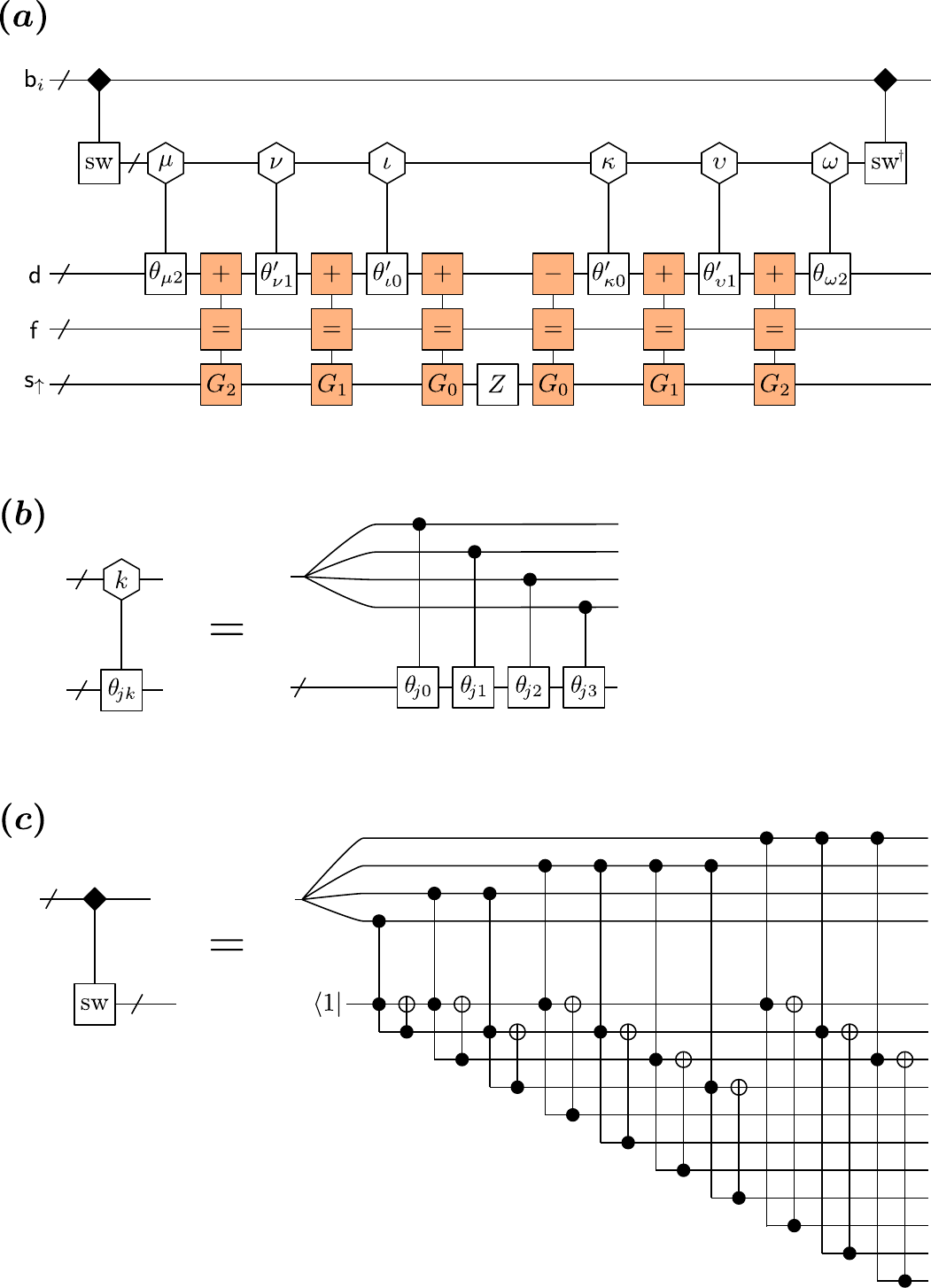}
    \caption{Unary encoding in the $\mathsf{Select}$ circuit. $\boldsymbol{(a)}$ Selecting the operators $U_{\mu} Z_{0} U_{\mu}^\dagger$ with a unary circuit encoding the index $\mu$. This depiction uses the maximum amount of batching, here primed angles $\theta^\prime_{jk}$ are chosen such that they write that data $\theta_{jk}$ into a register already dirtied with the values $|\theta_{j\, (k-1)}\rangle$. At the beginning, the binary numbers held in the register $\mathsf{b}_i$ are converted into unary numbers. The register $\mathsf{d}$ is an additional resource, storing the angle. The gates used in this circuit are outlined in the next panels. $\boldsymbol{(b)}$ Unary version of a multiplexor. Here the data strings $\theta_{jk}$ are written into the lower register when the $k$-th qubits of the upper register are set. This circuit is comprised only of Clifford gates. $\boldsymbol{(c)}$ Out-of-place binary-to-unary conversion using left elbows. Performs the operation $|k\rangle |0\rangle \mapsto |k\rangle |2^{k+1}\rangle$, meaning the $(k+1)$-th qubit of the second register is in $|1\rangle$, while the others are in $|0\rangle$. This circuit is very similar to an inverse QRAM: the topmost qubit has been set to $|1\rangle$, and conditional swap operations move it to the respective position. Using the fact that the lower register starts in the collective $|0\rangle$ state, one can replace the conditional swaps with left elbows and Cliffords.}
    \label{fig:unary}
\end{figure}

\begin{figure}[tb]
    \centering
    \includegraphics[width=0.7\linewidth]{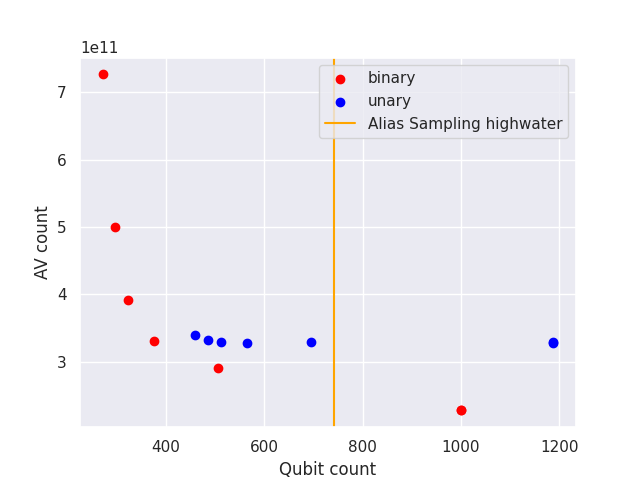}
     \caption{Active Volume and logic qubit count of the entire algorithm with respect to various degrees of batching, and distinguished by the version of $\mathsf{Select}$ that it uses. Instances in which the binary version of $\mathsf{Select}$, is used are denoted by red dots, while the blue data points represent instances in which the $\mathsf{\mathsf{b}}_i$ registers have been converted to unary. As an additional modification, we have turned the QROAM dataloader in the Alias Sampling procedure of $\mathsf{Prepare}$ back into a QROM. In this plot, instances of the binary $\mathsf{Select}$ seem much more preferable than the unary $\mathsf{Select}$ -- the blue datapoints seem to lie above the curve along which we find the red datapoints. This is happening not for the high AV of the unary circuit but for its overhead of $M+N/2$ qubits.}
    \label{fig:avbatchingunary}
\end{figure}

A modification of $\mathsf{Select}$ sees some internal data be converted from binary to unary. In the registers $\mathsf{b}_0$ and $\mathsf{b}_1$, the numbers $\mu$ are encoded as binary numbers, i.e. with $|\mu\rangle$ we mean 
$$ |\mu\rangle= |\mu_{\varkappa-1}\rangle \otimes  \cdots \otimes |\mu_{1}\rangle \otimes |\mu_{0}\rangle \, ,$$
for a register with $\varkappa$ qubits, and with bit values $\mu_x \in \lbrace0, 1 \rbrace$ such that $\mu = \sum_{x=0}^{\varkappa-1} \mu_x 2^x$. There is a quantum circuit that, considering a fresh register with $2^\kappa + 1$ qubits in $|0\rangle$ achieves an out-of-place conversion of $\mu$ into unary: $|\mu\rangle |0\rangle \mapsto |\mu\rangle |2^{\mu+1}\rangle$, where $|2^{\mu + 1}\rangle$ is a bit string where only the $(\mu+1)$-th (least significant) qubit is in $|1\rangle$, the other qubits are in $|0\rangle$, i.e. $|00010\rangle$ for $\mu = 1$. In $\mathsf{Select}$, we can use the binary-to-unary conversion for the registers $\mathsf{b}_i$: shifting the dataloaders from a binary $\mathsf{b}_i$ register to the respective entangled unary register allows us to nullify the circuit's magic state cost. This allows us to combine the binary-to-unary conversion with batching, as depicted in Figure \ref{fig:unary}. The caveat is that the binary-to-unary conversion circuit has roughly the magic state cost of one dataloader. However, with a maximum of $2^\varkappa + 1 = M + N/2 = 218$ qubits, hundreds of auxiliary qubits can be saved with batching. A $\mathsf{subselect}$ circuit, along with binary-to-unary conversion circuits and unary dataloading, is depicted in Figure \ref{fig:unary}. 
Unfortunately, it turns out that, in the AV world, the unary encoding version of $\mathsf{Select}$ is not more effective than $\mathsf{Select}$ of BLISS-THC-b, as is illustrated in Figure \ref{fig:avbatchingunary}. While the AV costs of the algorithm are decent, the unary datapoints in the figure (corresponding to various batch sizes after the unary conversion) lie strictly above the curve defined by the BLISS-THC-b algorithm. If unary did not have the overhead of $M+N/2$ qubits, the points would be shifted past the curve to the left. Perhaps, future work will address these shortcomings by hybridizing unary and binary encodings.
\section{Block Encoding of the Hamiltonian}
\label{sec:blencoding}
\begin{figure}
    \centering
    \includegraphics[width=0.5\linewidth]{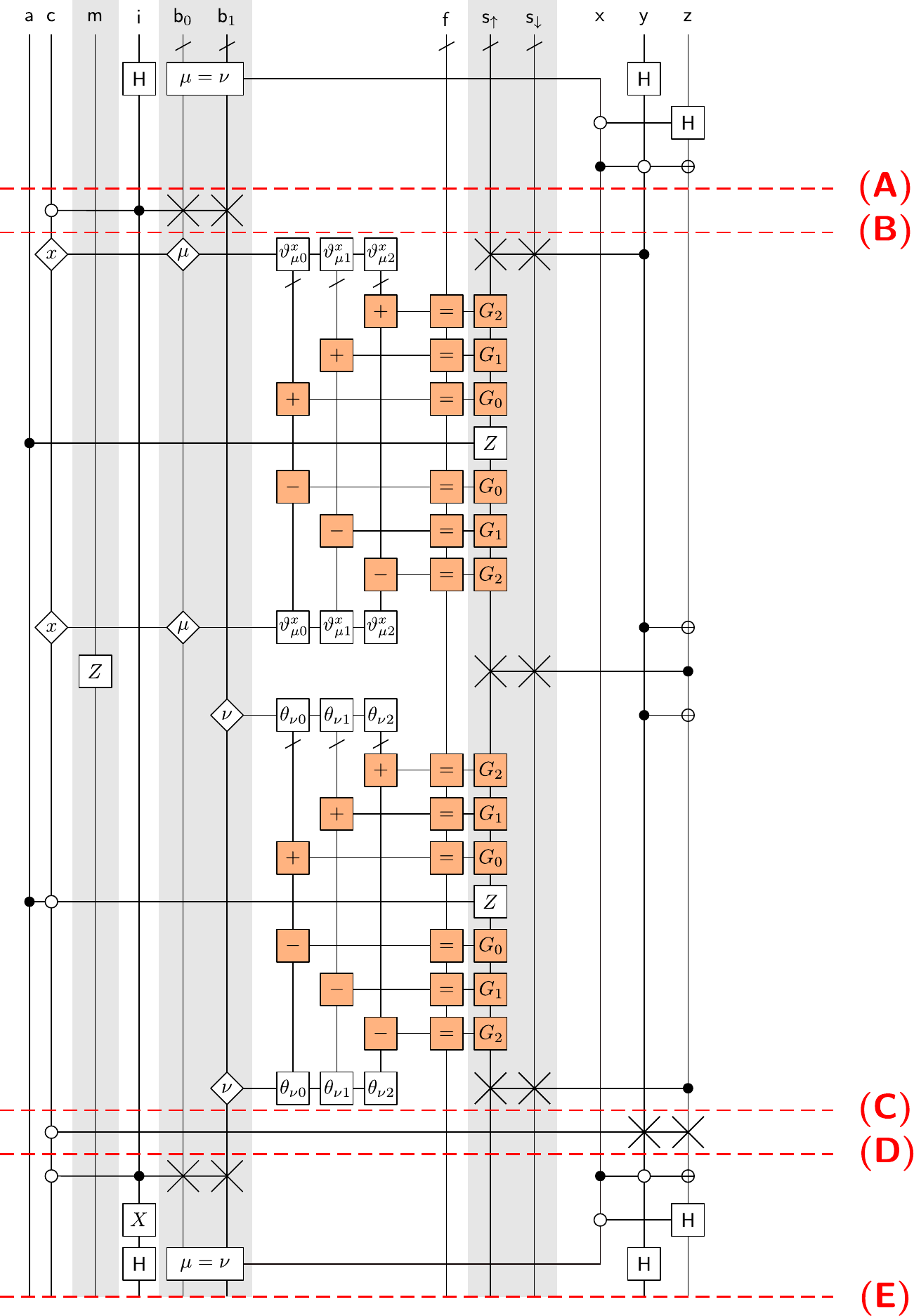}
    \caption{$\mathsf{Select}$ in Figure \ref{fig:select} divided into sections with the waypoints $(\mathsf{A})$, $(\mathsf{B})$, $(\mathsf{C})$, $(\mathsf{D})$ and $(\mathsf{E})$.}
    \label{fig:steps}
\end{figure}

In this section, we prove the block encoding property in Eq.~\eqref{eq:blocking}, using the $\mathsf{Select}$ circuit of Figure~\ref{fig:select} and justify the choice for the coefficients $\widehat{\zeta}_{\mu\nu}$ in Eq.~\eqref{eq:coeffs}. Let us consider the state
\begin{align}
|\Lambda_1\rangle = |1\rangle_{\mathsf{a}}\sum_{\nu=0}^M \sum_{\mu=0}^{\nu} \sqrt{\left| \widehat{\zeta}_{\mu\nu}\right|}\, |\mu\rangle_{\mathsf{b}_0}|\nu\rangle_{\mathsf{b}_1} |0\rangle_{\mathsf{i}}|\delta_{M,\nu}\rangle_{\mathsf{c}} |\mathrm{sign}(\widehat{\zeta}_{\mu\nu})\rangle_{\mathsf{m}} |0\rangle_{\mathsf{x}}|0\rangle_{\mathsf{y}}|0\rangle_{\mathsf{z}}\, ,
\end{align}
 a $p=1$ version of the state $|\Lambda\rangle$ in Eq.~\eqref{aux_state}. We now compare the Hamiltonian $H^\prime = \langle \Lambda_1| \mathsf{Select} |\Lambda_1\rangle $, encoded in the quantum computer, to the Hamiltonian to $\widetilde{H}$ of Eq.~\eqref{eq:biham}. To this end, we compute $\mathsf{Select}|\Lambda\rangle$ in 4 stages, represented by the states at the waypoints $(\mathsf{A})$-($\mathsf{E}$) in Figure~\ref{fig:steps}. In the first stage, we have apply a Hadamard gate to qubit $\mathsf{i}$ and have compute the qubits $\mathsf{x}$, $\mathsf{y}$ as well as $\mathsf{z}$, thus separating the cases $\nu=\mu$. Omitting $|1\rangle_{\mathsf{a}}$ from here on, $|\Lambda_1\rangle$ has become
\begin{align}\notag
 (\mathsf{A}):\quad &\sum_{\nu=0}^M \sum_{\mu=0}^{\nu-1} \sqrt{\left| \widehat{\zeta}_{\mu\nu}\right|}\, |\mu\rangle_{\mathsf{b}_0}|\nu\rangle_{\mathsf{b}_1} |\delta_{M,\nu}\rangle_{\mathsf{c}} |\mathrm{sign}(\widehat{\zeta}_{\mu\nu})\rangle_{\mathsf{m}}  |0\rangle_{\mathsf{x}}|+\rangle_{\mathsf{y}}|+\rangle_{\mathsf{z}}\\
&+  \frac{1}{\sqrt{2}}\sum_{\nu=0}^M  \sqrt{\left| \widehat{\zeta}_{\nu\nu}\right|}\, |\nu\rangle_{\mathsf{b}_0}|\nu\rangle_{\mathsf{b}_1} |\delta_{M,\nu}\rangle_{\mathsf{c}} |\mathrm{sign}(\widehat{\zeta}_{\mu\nu})\rangle_{\mathsf{m}}  |1\rangle_{\mathsf{x}} \left( |0\rangle_{\mathsf{y}}|1\rangle_{\mathsf{z}} + |1\rangle_{\mathsf{y}}|0\rangle_{\mathsf{z}}\right)\, ,
\end{align}
at waypoint $(\mathsf{A})$.
In the next stage, the registers $\mathsf{b}_0$ and $\mathsf{b}_1$ are swapped, but only in the subspace reserved for 2-body operators: the $|0\rangle$ subspace of $\mathsf{c}$. At waypoint $(\mathsf{B})$ the state becomes
\begin{align}
(\mathsf{B}):\quad \frac{1}{\sqrt{2}}&\sum_{\nu=0}^{M-1} \sum_{\mu=0}^{\nu-1} \sqrt{\left| \widehat{\zeta}_{\mu\nu}\right|}\, \left(\vphantom{\frac{1}{1}} |\mu\rangle_{\mathsf{b}_0}|\nu\rangle_{\mathsf{b}_1}|0\rangle_\mathsf{i}  + |\nu\rangle_{\mathsf{b}_0}|\mu\rangle_{\mathsf{b}_1}|1\rangle_\mathsf{i} \right)|0\rangle_{\mathsf{c}} |\mathrm{sign}(\widehat{\zeta}_{\mu\nu})\rangle_{\mathsf{m}}  |0\rangle_{\mathsf{x}}|+\rangle_{\mathsf{y}}|+\rangle_{\mathsf{z}} \notag \\
&+  \frac{1}{\sqrt{2}}\sum_{\nu=0}^M  \sqrt{\left| \widehat{\zeta}_{\nu\nu}\right|}\, |\nu\rangle_{\mathsf{b}_0}|\nu\rangle_{\mathsf{b}_1} |+\rangle_\mathsf{i}|\delta_{M,\nu}\rangle_{\mathsf{c}} |\mathrm{sign}(\widehat{\zeta}_{\mu\nu})\rangle_{\mathsf{m}}  |1\rangle_{\mathsf{x}} \left(\vphantom{\frac{1}{1}} |0\rangle_{\mathsf{y}}|1\rangle_{\mathsf{z}} + |1\rangle_{\mathsf{y}}|0\rangle_{\mathsf{z}}\right)\notag\\
&+ \sum_{\mu=0}^{N/2-1} \sqrt{\left| \widehat{\zeta}_{\mu M}\right|}\, |\mu\rangle_{\mathsf{b}_0}|M\rangle_{\mathsf{b}_1}|+\rangle_\mathsf{i} |1\rangle_{\mathsf{c}} |\mathrm{sign}(\widehat{\zeta}_{\mu M})\rangle_{\mathsf{m}}  |0\rangle_{\mathsf{x}}|+\rangle_{\mathsf{y}}|+\rangle_{\mathsf{z}}
\end{align}
The following stage applies a $Z$ operator on the sign qubit $\mathsf{m}$, as well as the (modified) $\mathsf{subselect}$ operators
\begin{align}
\sum_\sigma\sum_\mu |\mu\rangle\!\langle\mu|_{\mathsf{b}_0}\otimes |\sigma\rangle\!\langle\sigma|_{\mathsf{y}}\otimes \left(|0\rangle \langle 0|_{\mathsf{c}} \otimes  \widetilde{Z}_{\mu,\sigma} +   |1\rangle \langle 1|_{\mathsf{c}} \otimes  \widetilde{T}_{\mu,\sigma}\right)
\end{align}
and
\begin{align}
\sum_\tau\sum_\nu |\nu\rangle\!\langle\nu|_{\mathsf{b}_1}\otimes |\tau\rangle\!\langle\tau|_{\mathsf{z}} \otimes  \widetilde{Z}_{\nu,\tau} \, ,
\end{align}
where
\begin{align}
    \widetilde{Z}_{\mu,\sigma} = U^{{\dagger}}_{\mu} Z_{0,\sigma} U^{\phantom{\dagger}}_{\mu} \qquad \text{and} \qquad     \widetilde{T}_{\mu,\sigma} = V^{{\dagger}}_{\mu} Z_{0,\sigma} V^{\phantom{\dagger}}_{\mu} \, .
\end{align}
At waypoint $(\mathsf{C})$, the state thus becomes
\begin{align}
(\mathsf{C}):\quad & \frac{1}{\sqrt{8}}\sum_{\nu=0}^{M-1} \sum_{\mu=0}^{\nu-1} \sqrt{\left| \widehat{\zeta}_{\mu\nu}\right|}\, \sum_{\sigma\tau} \left(\vphantom{\frac{1}{1}} |\mu\rangle_{\mathsf{b}_0}|\nu\rangle_{\mathsf{b}_1}|0\rangle_\mathsf{i}\widetilde{Z}_{\nu,\tau}\widetilde{Z}_{\mu,\sigma}  + |\nu\rangle_{\mathsf{b}_0}|\mu\rangle_{\mathsf{b}_1}|1\rangle_\mathsf{i} \widetilde{Z}_{\mu,\tau}\widetilde{Z}_{\nu,\sigma} \right)|0\rangle_{\mathsf{c}} Z_{\mathsf{m}} |\mathrm{sign}(\widehat{\zeta}_{\mu\nu})\rangle_{\mathsf{m}}  |0\rangle_{\mathsf{x}}|\sigma\rangle_{\mathsf{y}}|\tau\rangle_{\mathsf{z}}\notag \\
& + \frac{1}{\sqrt{2}}\sum_{\nu=0}^{M-1}  \sqrt{\left| \widehat{\zeta}_{\nu\nu}\right|}\, |\nu\rangle_{\mathsf{b}_0}|\nu\rangle_{\mathsf{b}_1} |+\rangle_\mathsf{i}|0\rangle_{\mathsf{c}} Z_{\mathsf{m}}|\mathrm{sign}(\widehat{\zeta}_{\mu\nu})\rangle_{\mathsf{m}}  |1\rangle_{\mathsf{x}} \left(\vphantom{\frac{1}{1}} |0\rangle_{\mathsf{y}}|1\rangle_{\mathsf{z}} \widetilde{Z}_{\nu,\downarrow}\widetilde{Z}_{\nu,\uparrow} + |1\rangle_{\mathsf{y}}|0\rangle_{\mathsf{z}} \widetilde{Z}_{\nu,\uparrow}\widetilde{Z}_{\nu,\downarrow}\right)\notag\\
& + \frac{1}{\sqrt{2}} \sum_{\mu=0}^{N/2-1} \sqrt{\left| \widehat{\zeta}_{\mu M}\right|}\, |\mu\rangle_{\mathsf{b}_0}|M\rangle_{\mathsf{b}_1}|+\rangle_\mathsf{i} |1\rangle_{\mathsf{c}} Z_{\mathsf{m}}|\mathrm{sign}(\widehat{\zeta}_{\mu M})\rangle_{\mathsf{m}}  |0\rangle_{\mathsf{x}}\left(\vphantom{\frac{1}{1}} |0\rangle_{\mathsf{y}} \widetilde{T}_{\mu,\uparrow} +  |1\rangle_{\mathsf{y}} \widetilde{T}_{\mu,\downarrow}\right)|+\rangle_{\mathsf{z}}\, .
\end{align}
The gate in the next stage swaps the qubits $\mathsf{y}$ and $\mathsf{z}$ in the $|0\rangle$ subspace of $\mathsf{c}$. 
\begin{align}
(\mathsf{D}):\quad & \frac{1}{\sqrt{8}}\sum_{\nu=0}^{M-1} \sum_{\mu=0}^{\nu-1} \sqrt{\left| \widehat{\zeta}_{\mu\nu}\right|}\, \sum_{\sigma\tau} \left(\vphantom{\frac{1}{1}} |\mu\rangle_{\mathsf{b}_0}|\nu\rangle_{\mathsf{b}_1}|0\rangle_\mathsf{i}\widetilde{Z}_{\nu,\tau}\widetilde{Z}_{\mu,\sigma}  + |\nu\rangle_{\mathsf{b}_0}|\mu\rangle_{\mathsf{b}_1}|1\rangle_\mathsf{i} \widetilde{Z}_{\mu,\tau}\widetilde{Z}_{\nu,\sigma} \right)|0\rangle_{\mathsf{c}} Z_{\mathsf{m}} |\mathrm{sign}(\widehat{\zeta}_{\mu\nu})\rangle_{\mathsf{m}}  |0\rangle_{\mathsf{x}}|\tau\rangle_{\mathsf{y}}|\sigma\rangle_{\mathsf{z}}\notag \\
& + \frac{1}{\sqrt{2}}\sum_{\nu=0}^{M-1}  \sqrt{\left| \widehat{\zeta}_{\nu\nu}\right|}\, |\nu\rangle_{\mathsf{b}_0}|\nu\rangle_{\mathsf{b}_1} |+\rangle_\mathsf{i}|0\rangle_{\mathsf{c}} Z_{\mathsf{m}}|\mathrm{sign}(\widehat{\zeta}_{\mu\nu})\rangle_{\mathsf{m}}  |1\rangle_{\mathsf{x}} \left(\vphantom{\frac{1}{1}} |1\rangle_{\mathsf{y}}|0\rangle_{\mathsf{z}} \widetilde{Z}_{\nu,\downarrow}\widetilde{Z}_{\nu,\uparrow} + |0\rangle_{\mathsf{y}}|1\rangle_{\mathsf{z}} \widetilde{Z}_{\nu,\uparrow}\widetilde{Z}_{\nu,\downarrow}\right)\notag\\
& + \frac{1}{\sqrt{2}} \sum_{\mu=0}^{N/2-1} \sqrt{\left| \widehat{\zeta}_{\mu M}\right|}\, |\mu\rangle_{\mathsf{b}_0}|M\rangle_{\mathsf{b}_1}|+\rangle_\mathsf{i} |1\rangle_{\mathsf{c}} Z_{\mathsf{m}}|\mathrm{sign}(\widehat{\zeta}_{\mu M})\rangle_{\mathsf{m}}  |0\rangle_{\mathsf{x}}\left(\vphantom{\frac{1}{1}} |0\rangle_{\mathsf{y}} \widetilde{T}_{\mu,\uparrow} +  |1\rangle_{\mathsf{y}} \widetilde{T}_{\mu,\downarrow}\right)|+\rangle_{\mathsf{z}}\, .
\end{align}
In the last stage, we undo the swap of the second stage, flip qubit $\mathsf{i}$ and uncompute $\mathsf{i}$, $\mathsf{x}$, $\mathsf{y}$ and $\mathsf{z}$. 
Using  $\mathsf{H}|x\rangle =2^{-1/2}\sum_y (-1)^{xy}|y\rangle$ for bits $x,y \in \lbrace 0,1 \rbrace$, the state at waypoint $(\mathsf{E})$ is 
\begin{align}
(\mathsf{E}):\quad &  \frac{1}{8}\sum_{iyz}\sum_{\nu=0}^{M-1} \sum_{\mu=0}^{\nu-1} \sqrt{\left| \widehat{\zeta}_{\mu\nu}\right|}\, |\mu\rangle_{\mathsf{b}_0}|\nu\rangle_{\mathsf{b}_1} |i\rangle_\mathsf{i} \sum_{\sigma\tau} \left(\vphantom{\frac{1}{1}} (-1)^i \widetilde{Z}_{\nu,\tau}\widetilde{Z}_{\mu,\sigma}  +  \widetilde{Z}_{\mu,\tau}\widetilde{Z}_{\nu,\sigma} \right)|0\rangle_{\mathsf{c}} Z_{\mathsf{m}} |\mathrm{sign}(\widehat{\zeta}_{\mu\nu})\rangle_{\mathsf{m}}  |0\rangle_{\mathsf{x}}(-1)^{y\tau + z\sigma}|y\rangle_{\mathsf{y}}|z\rangle_{\mathsf{z}} \notag \\
& +\frac{1}{2}\sum_{y}\sum_{\nu=0}^{M-1}  \sqrt{\left| \widehat{\zeta}_{\nu\nu}\right|}\, |\nu\rangle_{\mathsf{b}_0}|\nu\rangle_{\mathsf{b}_1} |0\rangle_\mathsf{i}|0\rangle_{\mathsf{c}} Z_{\mathsf{m}}|\mathrm{sign}(\widehat{\zeta}_{\mu\nu})\rangle_{\mathsf{m}}  \left(\vphantom{\frac{1}{1}} (-1)^{y} \widetilde{Z}_{\nu,\downarrow}\widetilde{Z}_{\nu,\uparrow} + \widetilde{Z}_{\nu,\uparrow}\widetilde{Z}_{\nu,\downarrow}\right)|0\rangle_{\mathsf{x}} |y\rangle_{\mathsf{y}}|0\rangle_{\mathsf{z}} \notag \\
& + \frac{1}{2} \sum_y\sum_{\mu=0}^{N/2-1} \sqrt{\left| \widehat{\zeta}_{\mu M}\right|}\, |\mu\rangle_{\mathsf{b}_0}|M\rangle_{\mathsf{b}_1}|0\rangle_\mathsf{i} |1\rangle_{\mathsf{c}} Z_{\mathsf{m}}|\mathrm{sign}(\widehat{\zeta}_{\mu M})\rangle_{\mathsf{m}}  \left(\vphantom{\frac{1}{1}}  \widetilde{T}_{\mu,\uparrow} +   (-1)^y\, \widetilde{T}_{\mu,\downarrow}\right)|0\rangle_{\mathsf{x}}|y\rangle_{\mathsf{y}}|0\rangle_{\mathsf{z}}
\end{align}
Multiplying this state from the left-hand side with $\langle\Lambda_1|\cdot$ finally gives us 
\begin{align*}
    H^\prime = \frac{1}{8}\sum_{\mu=0}^{M-1} \sum_{\nu\neq \mu} \widehat{\zeta}_{\mu \nu} \sum_{\sigma\tau} \widetilde{Z}_{\nu,\tau} \widetilde{Z}_{\mu,\sigma} + \frac{1}{2}\sum_{\nu=0}^{M-1} \widehat{\zeta}_{\nu \nu}\sum_{\sigma} \widetilde{Z}_{\nu,\sigma} \widetilde{Z}_{\nu,\overline{\sigma}} + \frac{1}{2}\sum_{k=0}^{N/2-1} \widehat{\zeta}_{k M} \, \sum_{\sigma} \widetilde{T}_{k,\sigma} \, ,
\end{align*}
where $\overline{\sigma}$ the flipped version of $\sigma$. This is the Hamiltonian we have encoded, while the Hamiltonian we were aiming to encode is
\begin{align}
    \widetilde{H} = \frac{1}{8}\sum_{\mu=0}^{M-1} \sum_{\nu\neq \mu} \tilde{\zeta}_{\mu \nu} \sum_{\sigma\tau} \widetilde{Z}_{\nu,\tau} \widetilde{Z}_{\mu,\sigma} + \frac{1}{8}\sum_{\nu=0}^{M-1} \tilde{\zeta}_{\nu \nu}\sum_{\sigma} \widetilde{Z}_{\nu,\sigma} \widetilde{Z}_{\nu,\overline{\sigma}} - \frac{1}{2}\sum_{k=0}^{N/2-1} \tilde{t}_{k} \, \sum_{\sigma} \widetilde{T}_{k,\sigma} \, ,
\end{align}
where constant terms were eliminated with respect to Eq.~\eqref{eq:biham}.
We thus choose $\widehat{\zeta}_{\mu\nu}$ as in Eq.~\eqref{eq:coeffs}.
\end{document}